\pdfoutput=1
\documentclass[twoside,twocolumn,9pt]{article}
\usepackage{extsizes}
\usepackage[super,sort&compress,comma]{natbib} 
\usepackage[version=3]{mhchem}
\usepackage[left=1.5cm, right=1.5cm, top=1.785cm, bottom=2.0cm]{geometry}
\usepackage{balance}
\usepackage{times,mathptmx}
\usepackage{sectsty}
\usepackage{graphicx} 
\usepackage{lastpage}
\usepackage[format=plain,justification=justified,singlelinecheck=false,font={stretch=1.125,small,sf},labelfont=bf,labelsep=space]{caption}
\usepackage{float}
\usepackage{fancyhdr}
\usepackage{fnpos}
\usepackage[english]{babel}
\addto{\captionsenglish}{%
}
\usepackage{array}
\usepackage{droidsans}
\usepackage{charter}
\usepackage[T1]{fontenc}
\usepackage[usenames,dvipsnames]{xcolor}
\usepackage{setspace}
\usepackage[compact]{titlesec}
\usepackage{hyperref}

\usepackage{multicol,lipsum}

\usepackage{epstopdf}

\definecolor{cream}{RGB}{222,217,201}

\begin{document}

\pagestyle{fancy}
\thispagestyle{plain}
\fancypagestyle{plain}{

}

\makeFNbottom
\makeatletter
\renewcommand\LARGE{\@setfontsize\LARGE{15pt}{17}}
\renewcommand\Large{\@setfontsize\Large{12pt}{14}}
\renewcommand\large{\@setfontsize\large{10pt}{12}}
\renewcommand\footnotesize{\@setfontsize\footnotesize{7pt}{10}}
\makeatother

\renewcommand{\thefootnote}{\fnsymbol{footnote}}
\renewcommand\footnoterule{\vspace*{1pt}%
\color{cream}\hrule width 3.5in height 0.4pt \color{black}\vspace*{5pt}} 
\setcounter{secnumdepth}{5}

\makeatletter 
\renewcommand\@biblabel[1]{#1}            
\renewcommand\@makefntext[1]%
{\noindent\makebox[0pt][r]{\@thefnmark\,}#1}
\makeatother 
\renewcommand{\figurename}{\small{Fig.}~}
\sectionfont{\sffamily\Large}
\subsectionfont{\normalsize}
\subsubsectionfont{\bf}
\setstretch{1.125} 
\setlength{\skip\footins}{0.8cm}
\setlength{\footnotesep}{0.25cm}
\setlength{\jot}{10pt}
\titlespacing*{\section}{0pt}{4pt}{4pt}
\titlespacing*{\subsection}{0pt}{15pt}{1pt}

\fancyfoot{}
\fancyfoot[RO]{\footnotesize{\sffamily{1--\pageref{LastPage} ~\textbar  \hspace{2pt}\thepage}}}
\fancyfoot[LE]{\footnotesize{\sffamily{\thepage~\textbar\hspace{3.45cm} 1--\pageref{LastPage}}}}
\fancyhead{}
\renewcommand{\headrulewidth}{0pt} 
\renewcommand{\footrulewidth}{0pt}
\setlength{\arrayrulewidth}{1pt}
\setlength{\columnsep}{6.5mm}
\setlength\bibsep{1pt}

\makeatletter 
\newlength{\figrulesep} 
\setlength{\figrulesep}{0.5\textfloatsep} 

\newcommand{\topfigrule}{\vspace*{-1pt}%
\noindent{\color{cream}\rule[-\figrulesep]{\columnwidth}{1.5pt}} }

\newcommand{\botfigrule}{\vspace*{-2pt}%
\noindent{\color{cream}\rule[\figrulesep]{\columnwidth}{1.5pt}} }

\newcommand{\dblfigrule}{\vspace*{-1pt}%
\noindent{\color{cream}\rule[-\figrulesep]{\textwidth}{1.5pt}} }

\makeatother

\newcommand{\sta}{\mathrm{state}~\mathcal{A}} 
\newcommand{\stb}{\mathrm{state}~\mathcal{B}} 
\newcommand{\zcom}{x_{\rm COM}} 
\newcommand{\esw}{\epsilon_{\rm SW}} 

\newcommand{\Nt}{\tilde{N}} 
\newcommand{\Ntv}{\tilde{N}_v} 
\newcommand{\pvn}{P_v(N)}
\newcommand{\pvnt}{P_v(\tilde{N})}
\newcommand{\fvn}{F_v(N)}
\newcommand{\fvnt}{F_v(\tilde{N})}
\newcommand{\kbt}{k_{\rm{B}} T}
\newcommand{\Rbar}{\overline{\textbf{R}}}

\newcommand{\gsv}{\gamma_{\rm SV}} 
\newcommand{\gsl}{\gamma_{\rm SL}} 
\newcommand{\gvl}{\gamma_{\rm VL}} 

\newcommand{\SI}{\text{Supplementary Information}} 

\newcommand{\tcr}{\textcolor{red}}
\newcommand{\tcb}{\textcolor{blue}}
\newcommand{\tcg}{\textcolor{green}}

\newcommand{\cob}{\color{blue}} \newcommand{\com}{\color{magenta}}\newcommand{\cog}{\color{green} }
\renewcommand{\thesection}{\Alph{section}.} 
\renewcommand{\thesubsection}{\arabic{subsection}.}
%


\noindent\LARGE{\textbf{Characterizing Surface Wetting and Interfacial Properties using Enhanced Sampling (SWIPES)$^\dag$}} \\
\vspace{0.3cm} 

\noindent\large{
Hao Jiang, \textit{$^{a}$} 
Suruchi Fialoke, \textit{$^{a}$} 
Zachariah Vicars, \textit{$^{a}$} and 
Amish J. Patel \textit{$^{a}$}
} \\


\begin{center}
\textbf{Abstract}
\end{center}
We introduce an accurate and efficient method for characterizing surface wetting and interfacial properties, such as
the contact angle made by a liquid droplet on a solid surface, and the vapor-liquid surface tension of a fluid.
The method makes use of molecular simulations in conjunction with the indirect umbrella sampling technique 
to systematically wet the surface and estimate the corresponding free energy.
To illustrate the method, we study the wetting of a family of Lennard-Jones surfaces by water. 
We estimate contact angles for surfaces with a wide range of attractions for water by using our method 
and also by using droplet shapes.
Notably, as surface -- water attractions are increased, our method is able to capture the transition from partial to complete wetting. 
Finally, the method is straightforward to implement and computationally efficient, 
providing accurate contact angle estimates in roughly 5 nanoseconds of simulation time. \\





\renewcommand*\rmdefault{bch}\normalfont\upshape
\rmfamily
\section*{}
\vspace{-1cm}


\footnotetext{\textit{$^{a}$~
Department of Chemical and Biomolecular Engineering, University of Pennsylvania, Philadelphia, PA 19104, United States. 
E-mail: amish.patel@seas.upenn.edu}}
%

\footnotetext{\dag~Electronic Supplementary Information (ESI) available: 
included are 
details on the placement of the observation volume ($v$);
proof that interface positions ($H$) and free energies ($F_{\kappa,N^*}$) vary linearly with $N^*$;
illustration that $f$ and $h$ can be obtained from individual biased simulations;
dependences of $F_{\kappa,N^*}$ and $\langle \zcom \rangle_{\kappa,N^*}$ on $N^*$ for LJ surfaces with different $\esw$-values, 
and for different inter-surface separations;
vapor-liquid interface profiles obtained from our biased simulations; and
shapes of cylindrical water droplets in contact with the LJ surface.
}


\section{Introduction}
Wetting of solid surfaces by fluids is important in diverse disciplines, including but not limited to surface chemistry, materials characterization, oil and gas recovery~\cite{Mittal,deshmukh2012atomic, stark2013surface, hu2014microscopic, prakash2017quantifying}. 
%
In general, the wettability of a solid by a fluid is characterized by 
a wetting coefficient, $k \equiv (\gsv - \gsl ) / \gvl$, where $\gamma$ represents surface tension,
and the subscripts correspond to the coexisting vapor (V), liquid (L) and solid (S) phases.
The wetting coefficient is also related to the contact angle ($\theta$) that a liquid droplet (surrounded by its vapor) makes with a solid surface; according to Young's equation, $\cos\theta = (\gsv - \gsl)/\gvl = k$. 
Thus, the extent to which a fluid prefers to wet a solid,
or the preference of the solid for the liquid over its vapor, 
can be characterized by estimating either $k$ or $\theta$.

The most common approach for characterizing wettability is the direct measurement 
of contact angle using the so-called sessile droplet method, 
wherein $\theta$ is determined from the geometry of a liquid droplet supported by a solid surface. 
Although the sessile droplet method is usually associated with the experimental determination of $\theta$,
molecular simulations have also made extensive use of this method.
Indeed, both molecular dynamics (MD) and Monte Carlo (MC) simulations have been used to estimate contact angles for diverse interfacial systems, ranging from toy models~\cite{binder2003monte, Ingebrigtsen, Shi, Becker, Malani} to realistic molecular models such as water/minerals~\cite{Koishi, Tsuji, Goual, Chen, Cygan, Firoozabadi, Bejagam} and textured surfaces~\cite{Kumar:JCP:2011,Leroy:JCTC:2012,Shahraz:Langmuir:2013}.
However, in contrast with the millimeter-sized sessile droplets used in experiments, the droplets used in molecular simulations are nanoscopic.
At the nanoscale, the geometry of spherical droplets can be influenced by the tension associated with the three-phase contact line. 
As a result, estimates of $\theta$ can depend strongly on the size of the simulated droplet~\cite{Weijs, Peng}. 
Such line tension effects can be mitigated by using a cylindrical droplet that is infinitely long (due to periodic boundaries), 
because the three-phase contact line then has a fixed length and is independent of droplet size~\cite{slovin2015identifying}.
However, a number of studies have recently shown that $\theta$ extracted from the geometry of 
such cylindrical droplets nevertheless depends on the curvature of the droplet~\cite{Kanduc, Scocchi,Kanduc2018}. 
In addition to the challenges posed by line tension and finite size effects, the determination of $\theta$ from droplet geometry 
is also plagued by a certain degree of ambiguity regarding the exact location of the solid-fluid interface~\cite{Skvara,Ravipati,Khalkhali}.
%

Given the limitations of the sessile droplet method, 
a number of free energy methods have been proposed 
for estimating $k$.
%
The interface potential approach involves estimating the free energy per unit area needed: 
(i) to wet a surface in contact with vapor, $\gvl + \gsl - \gsv = \gvl(1- k)$ (spreading potential), and
(ii) to dewet a surface in contact with liquid, $\gvl + \gsv - \gsl = \gvl(1+ k)$ (drying potential),
thereby providing estimates of both $\gvl$ and $k$~\cite{Errington1}.
Errington and co-workers have illustrated the utility of this approach, and have employed it extensively 
in conjunction with grand canonical Monte Carlo simulations to study the wetting properties of diverse surfaces by a number of fluids~\cite{Errington1, Errington3,Errington5}.
However, because using grand canonical Monte Carlo to simulate large or complex fluid molecules can be inefficient,
and performing molecular dynamics simulations in the grand canonical ensemble is cumbersome,
the interface potential method has not been applied to the study of such complex fluids.
%
To this end, approaches that employ molecular dynamics simulations have been developed, e.g.,
the phantom-wall and dry-surface methods by Leroy and Muller-Plathe~\cite{Leroy, Leroy1,Leroy2}. 
These approaches are similar in spirit to the drying potential described above, i.e., they employ processes that enable estimation of the work of adhesion. 
In the phantom-wall method, a structureless repulsive wall is employed to push the liquid away from the surface of interest~\cite{Leroy, Leroy1}, whereas in the dry-surface method, the surface-water attractions are reversibly turned off~\cite{Leroy2}.
These methods have been widely applied to calculate wetting properties for a number of different systems~\cite{Leroy,Taherian,Jiang,Leroy4}.
However, in each of the above free energy methods, the solid-liquid interface is replaced by both a solid-vapor and a vapor-liquid interface~\cite{Errington1,Patel:PNAS:2011,Leroy, Leroy1}.
Such formation of vapor-liquid interfaces in the presence of periodic boundary conditions involves transitions between dewetted morphologies~\cite{Binder,Remsing:PNAS:2015,Prakash},
which can give rise to hysteresis and complicate estimation of the associated free energies~\cite{Godawat,macdowell2006nucleation,grzelak2010nanoscale}.
%

Moreover, in the phantom-wall method, a solid-liquid interface is replaced not by a solid-vapor interface, but by a solid-vacuum interface.
The corresponding free energy per unit area, $\gvl + \gamma_{\rm S,Vac} - \gsl$
approximates $\gvl + \gsv - \gsl = \gvl(1+k)$
if the solid-vacuum surface tension, $\gamma_{\rm S,Vac} \approx \gsv$.
Such an assumption is reasonable when solid-fluid attractions are relatively weak, 
and there little to no adsorption from the vapor phase onto the solid surface.
However, $\gamma_{\rm S,Vac}$ is expected to differ substantially from $\gsv$ when the solid-fluid attractions are significant. 
%
Indeed, Kanduc and Netz~\cite{Netz1, Netz2} have recently shown that for a sufficiently solvophilic surface, 
there is substantial adsorption from the vapor phase onto the solid surface, which lowers $\gsv$ (relative to $\gamma_{\rm S,Vac}$).
To do so, the authors introduced a free energy method that enables estimation of the spreading potential, 
and is able to capture the transition from partial to complete wetting as solid-fluid attractions are increased.
However, the method is computationally demanding and requires roughly 50 simulations for each estimate of $k$. 

Here, we introduce a versatile and computationally efficient method,
which we call ``Surface Wetting and Interfacial Properties using Enhanced Sampling'' or SWIPES.
SWIPES makes use of the indirect umbrella sampling (INDUS) technique~\cite{Patel:JPCB:2010,Patel}
to systematically wet the surface of interest, and estimate the free energy change associated with the process. 
It can be used in conjunction with molecular dynamics simulations to obtain estimates of 
not just wetting coefficients $k$, but also of vapor-liquid surface tensions, $\gvl$. 
We illustrate SWIPES by using it to study the wetting of Lennard-Jones (LJ) surfaces by SPC/E water~\cite{SPCE}. 
We estimate wetting coefficients for surfaces with a wide range of solid-water attractions,
and compare our results with those obtained from droplet geometries.
We also show that our method is capable of capturing the transition from partial to complete wetting as the solid-water affinity is increased.  
The rest of the paper is organized as follows: 
In section B, our approach for estimating wetting coefficients is described.
Details pertaining to the molecular models used and the simulation methods employed are given in section C.
In section D, contact angles calculated from free energies and from droplet geometries are compared,
and the transition from partial to complete wetting as well as the computational efficiency of the method are discussed. 
Finally, we summarize our findings in the conclusions section.

\begin{figure}[h]
   \centering
    \includegraphics[width=0.45\textwidth]{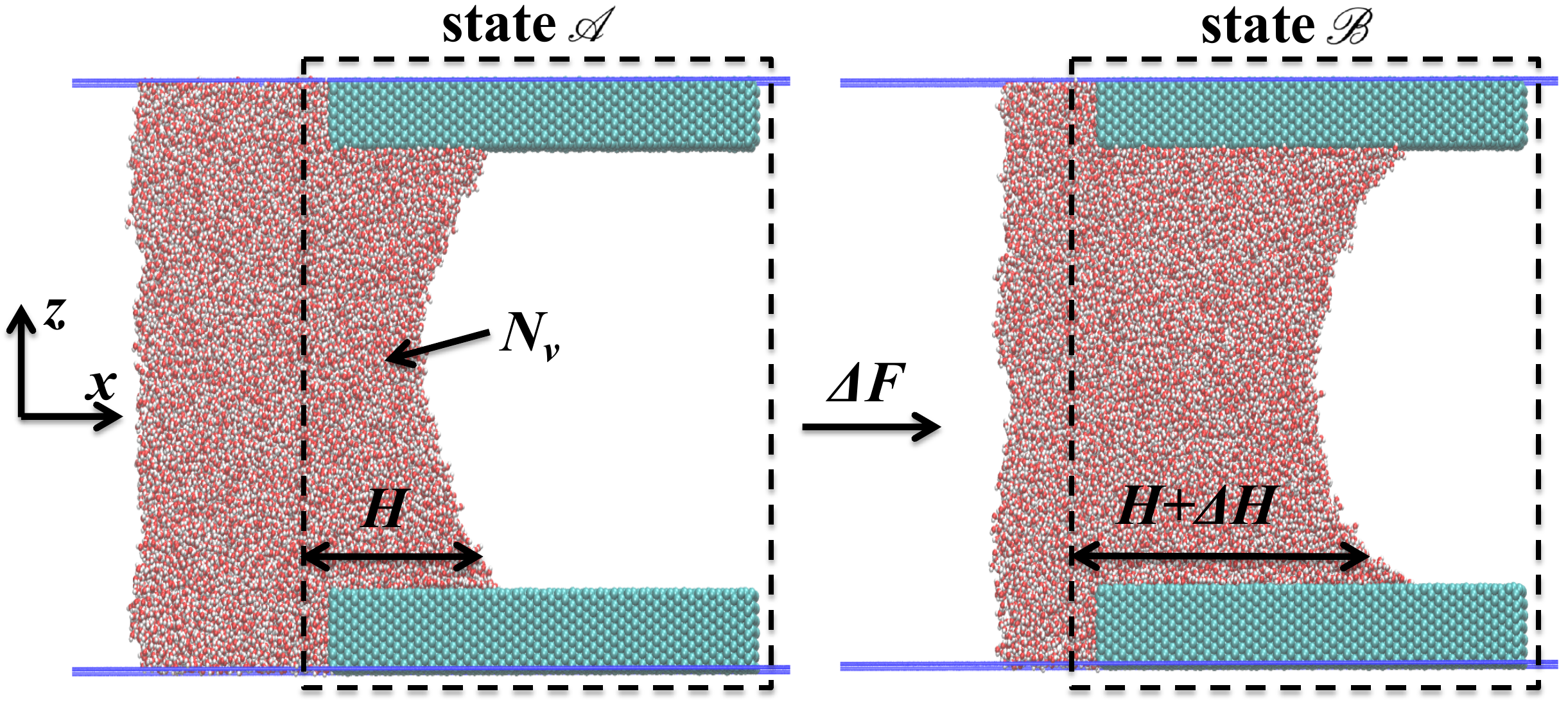} 
    \caption{
Schematic illustrating our approach. 
The solid surface of interest is represented using cyan spheres. 
The dashed line marks the observation volume, $v$. 
The number of fluid molecules (in this case water, shown in red and white) inside the observation volume is denoted by $N_v$. 
A biasing potential is used to modulate $N_v$,  
which results in the vapor-liquid interface 
moving along the solid surface by a distance $\Delta H$. 
The wetting coefficient ($k$) and the contact angle ($\theta$) can be obtained from estimates of $\Delta H$,
and the free energy change associated with this process, $\Delta F$; see Equation~\ref{eq:wil2}.
}
    \label{fig0} 
  \end{figure}

\vspace{-0.1in}
\section{Estimating Wetting Coefficients}
%
In this section, we describe our approach for calculating the wetting coefficient, $k$, defined as:
\begin{equation}
k \equiv \frac{\gsv - \gsl}{\gvl}
\label{eq:kdef}
\end{equation}
The wetting coefficient is related to the contact angle ($\theta$) through $k = \cos\theta$.  

\subsection{Outline of Our Approach}
%
To estimate $k$ for the wetting of a flat solid surface by a liquid in equilibrium with its vapor, we employ the schematic shown in Figure~\ref{fig0}.
As the number of waters, $N_v$, in the observation volume, $v$, increases, and the system moves from $\sta$ to $\stb$, 
the area of the solid wetted by the liquid increases by $\Delta A = 2 L \Delta H$, where $L$ is the length of the simulation box 
in the direction perpendicular to the page, and $\Delta H$ is the distance that the vapor-liquid interface advances along the solid.
The work, $\Delta F$, that must be done to wet the solid as the system moves from $\sta$ to $\stb$, is given by
\begin{equation}
\Delta F = (\gsl  - \gsv)\Delta A = -k \gvl \Delta A = -k \gvl \cdot 2 L \Delta H.
\label{eq:wil1}
\end{equation}
Although a (curved) vapor-liquid interface and three-phase contact lines are present in our setup,
the corresponding interfacial area, its curvature, and the contact line length are all unaffected 
by the advance of the liquid slab along the solid surface, and thereby do not influence $\Delta F$.
The wetting coefficient ($k$) or the contact angle ($\theta$) can thus be estimated from the ratio of $\Delta F$ and $\Delta H$ as
\begin{equation}
k = \cos\theta = -\frac{1}{2 \gvl L} \cdot \frac{\Delta F}{ \Delta H}.
\label{eq:wil2}
\end{equation}
We note that the ratio $\Delta F$/$\Delta H$ ought to be the same ($-2 \gvl L k$), 
regardless the positions of vapor-liquid interface in $\sta$ and $\stb$,
as long as the solid surface under consideration is homogeneous, i.e., free of chemical heterogeneity or roughness. 

\subsection{Indirectly Sampling $\Ntv$}
To efficiently estimate $\Delta F$/$\Delta H$, we employ enhanced sampling simulations that pull waters into $v$ and wet the surface.
In principle, this can be accomplished by biasing the number of waters, $N_v$, in the rectangular observation volume, $v$, shown by the dashed lines in Figure~\ref{fig0}.
%
In practice, because $N_v$ is a discrete function of particle positions, $\Rbar$, biasing it using molecular dynamics simulations would result in impulsive forces. 
We therefore bias $N_v$ indirectly by instead biasing a coarse-grained number of fluid molecules, $\Ntv$, following the indirect umbrella sampling (INDUS) prescription~\cite{Patel:JPCB:2010,Patel};
$\Ntv$ is closely related to $N_v$, but is a continuous function of $\Rbar$.
The precise definition of $\Ntv$ and its dependence on $\Rbar$ can be found in the ref.~\citenum{Patel}. 
%
Our choice of the location of $v$ is discussed in further detail in the $\SI$.
%

\subsection{Biased Simulations to Wet the Surface}
%
We choose a biasing potential that is parabolic in $\Ntv(\Rbar)$,  
\begin{equation}
U_{\kappa,N^*}(\Ntv) = \frac{\kappa}{2} ( \Ntv - N^* )^2,
\end{equation}
with $\kappa$ and $N^*$ parametrizing the potential.
We further choose $\sta$ and $\stb$ to correspond to biased ensembles with different values of $N^*$, say $N_A^*$ and $N_B^*$.
Although this choice is convenient, it is certainly not unique; alternative definitions can be chosen for the two states, e.g., constant-$\Ntv$ ensembles with different $\Ntv$-values.
Nevertheless, this choice enables us to reversibly wet the surface, and to estimate the corresponding free energetics,
which we characterize using the free energy difference between the biased and unbiased ensembles, $F_{\kappa,N^*}$.
Similarly, to characterize the position of the vapor-liquid interface, $H$, we employ the biased ensemble average 
of the center of mass of water in the simulation box in the direction ($x$) perpendicular to the vapor-liquid interface, i.e., 
$H \equiv \langle \zcom \rangle_{\kappa,N^*}$,
where $\langle \mathcal{O}(\Rbar) \rangle_{\kappa,N^*}$ represents the average of $\mathcal{O}(\Rbar)$ in the biased ensemble.
As shown in the $\SI$, 
both the position of the vapor-liquid interface, $H$, and the corresponding free energy, $F_{\kappa,N^*}$, vary linearly with $N^*$; 
i.e., both $f \equiv dF_{\kappa,N^*} / dN^*$ and $h \equiv dH / dN^*$ are independent of $N^*$.
Briefly, this stems from the fact that the balance of forces at the 3-phase contact line is governed entirely by the corresponding interfacial tensions,
and the shape of the vapor-liquid interface is independent of the biasing potential.
Thus, the ratio $\Delta F / \Delta H$ needed to obtain $k$ (using Equation~\ref{eq:wil2}) can be estimated as a ratio of the corresponding slopes, i.e.,
\begin{equation}
\frac{ \Delta F }{ \Delta H }  = \frac{ dF_{\kappa,N^*} / dN^* }{ dH / dN^* } = \frac{f}{h}
\end{equation}

\subsection{Estimating $f$ and $h$}
%
To estimate $f$, we use the thermodynamic integration formula,
\begin{equation}
f \equiv \frac{dF_{\kappa,N^*}}{dN^*} =  \bigg\langle \frac{ d U_{\kappa,N^*} }{ dN^* } \bigg\rangle_{\kappa,N^*} = \kappa [ N^* - \langle \Ntv \rangle_{\kappa,N^*} ]. 
\end{equation}
Thus, $f$ can be obtained using a single simulation through
\begin{equation}
f = - \kappa [ \langle \Ntv \rangle_{\kappa,N^*} - N^* ].
\label{eq:f}
\end{equation}
Alternatively, by rearranging Equation~\ref{eq:f}, we get
\begin{equation}
\langle \Ntv \rangle_{\kappa,N^*} = N^* - f / \kappa,
\label{eq:N}
\end{equation}
%
suggesting that $\langle \Ntv \rangle_{\kappa,N^*}$ ought to differ from $N^*$ by a constant offset, $-f / \kappa$.
Thus, an alternate way to estimate $f$ from multiple simulations (with the same $\kappa$ but different $N^*$-values) 
would be to plot $\langle \Ntv \rangle_{\kappa,N^*}$ versus $N^*$;
the data can then be fitted to a straight line with a unit slope, and the $y$-intercept used to estimate $-f / \kappa$.
This procedure is functionally equivalent to averaging estimates of $f$ obtained from different biased simulations using Equation~\ref{eq:f}.

Similarly, $h$ can be obtained from multiple biased simulations (with the same $\kappa$) by fitting 
$H \equiv \langle \zcom \rangle_{\kappa,N^*}$ versus $N^*$ to a straight line, and obtaining the slope.
Alternatively, $h$ can also be obtained from a single simulation using the co-variance relation
\begin{equation}
h \equiv \frac{dH }{ dN^*} = \frac{d \langle \zcom \rangle_{\kappa,N^*} } { dN^*} = \beta\kappa \langle \delta \zcom \delta N_v \rangle_{\kappa,N^*},
\label{eq:covar}
\end{equation}
where $\beta^{-1} \equiv k_{\rm B}T$, $k_{\rm B}$ is the Boltzmann constant and $T$ is the system temperature.
Once both $f$ and $h$ have been estimated, the product of the wetting coefficient and surface tension can be readily obtained as
\begin{equation}
k \gvl = -\frac{1}{2 L} \cdot \frac{f}{h}
\label{eq:kdef2}
\end{equation}
For a surface with a known value of $k$, Equation~\ref{eq:kdef2} can then be used to obtain the vapor-liquid surface tension, $\gvl$.
Similarly, for a liquid with a known $\gvl$, $k$ can be estimated using Equation~\ref{eq:kdef2}.

\subsection{Characteristic Features of SWIPES}
%
The setup used in SWIPES is similar in spirit to the Wilhelmy plate method, 
which is a commonly-used experimental technique for determining $\gvl \cos\theta$~\cite{Butt,alghunaim2016techniques}. 
In the experiments, the surface of interest (the Wilhelmy plate) is immersed in the liquid, and the capillary force it experiences is measured;
here we bias the fluid to wet the surface instead, and measure the corresponding free energy change per unit length ($f/h$).
%
The SWIPES approach is versatile, and can be used not only with Monte Carlo, but also with molecular dynamics simulations.
It can be used to estimate both $\gvl$ and $k$ for fluids and surfaces that vary widely in their cohesive and adhesive interaction strengths.
Moreover, SWIPES is relatively easy to implement, and because it only requires estimates of ensemble averages that converge quickly, it is also computationally efficient.
Another salient feature of SWIPES
is that as the number of fluid molecules in $v$ increases, 
the vapor-liquid interface moves along the solid surface, 
but no additional vapor-liquid interfacial area is created. 
In contrast, either vapor-liquid or vacuum-liquid interfaces are created in several computational approaches for estimating $k$, 
including the interface potential~\cite{Errington1,Errington3}, phantom-wall~\cite{Leroy1,Leroy}, and dry-surface methods~\cite{Leroy2}.
When carried out in the presence of periodic boundary conditions,
the creation of such interfaces involves transitions between different dewetted morphologies, 
which can give rise to hysteresis effects, and complicate estimation of the corresponding free energies~\cite{Godawat,macdowell2006nucleation,grzelak2010nanoscale,Netz1}.

%

\section{Molecular Models and Simulation Details}
To illustrate our approach for estimating $\gvl$ and $k$, we study the wetting of a Lennard-Jones (LJ) solid surface by SPC/E water~\cite{SPCE}.
Following ref.~\citenum{Leroy2}, we study a family of surfaces composed of LJ atoms that vary in the strength of their interactions with water.
The LJ atoms are placed on an FCC lattice with a spacing of 4.05 \AA, and are restrained to their initial positions using harmonic springs with a spring constant of 1000~kJ/mol/nm$^2$. 
The (111) face of the LJ surface is placed in contact with water.
The LJ interaction parameters for the solid atoms are $\sigma = 0.2629$~nm and $\epsilon = 22.13$~kJ/mol. 
The cross interaction between the solid atoms and water oxygens follows the Lorentz-Berthelot combining rule for $\sigma$, 
whereas the corresponding well-depth is systematically varied from $\esw = 0.001$ to $2.6$~kJ/mol to tune surface-water interaction strength.
The cutoff distance is chosen to be 1 nm for both the LJ potential and the real space electrostatic interactions.
The reciprocal space electrostatic interactions are treated using particle mesh Ewald summation with a Fourier spacing of 0.1 nm~\cite{Smit}. 
The equations of motion are integrated using the leap-frog algorithm with a time step of 2 fs.  
The simulations are performed in the canonical ensemble, and the temperature of the system is set to 300 K using a Langevin thermostat with a coupling constant of 2 ps~\cite{Schneider}. 
The SHAKE algorithm is used to constrain the internal degrees of freedom of the SPC/E water molecule~\cite{shake}. 
All simulations were performed using GROMACS MD simulation package~\cite{gromacs}, which was modified in-house to incorporate the biasing potentials of interest. 
The biasing potentials described here have also been implemented in the open-source INDUS code, which can be accessed at:
http://patelgroup.seas.upenn.edu/indus.html.
The INDUS code implements the INDirect Umbrella Sampling method for biasing coarse-grained particle number as an extension to the PLUMED plugin, which interfaces with several popular MD simulation packages~\cite{plumed}.

\subsection{Biased Simulations}
A solid surface consisting of 8640 particles is placed in contact with 7000 water molecules 
in a rectangular box with dimensions of 26.5 nm $\times$ 2.8059 nm $\times$ 14.319 nm (Figure~\ref{fig0}). 
Additionally, a repulsive wall (not shown in Figure~\ref{fig0}) consisting of four layers of LJ particles is placed 
at the left edge ($-x$) of the simulation box to break translational symmetry and nucleate vapor in its vicinity.
The wall has 1200 particles that interact with water oxygens through the LJ potential with $\sigma = 0.29$~nm and a small $\epsilon = 0.001$~kJ/mol.
As illustrated in Figure~\ref{fig0}, the left edge of the observation volume is placed roughly 1~nm to the left of the solid surface of interest, 
and the observation volume covers the entire surface.
To obtain estimates of $f$ and $h$, twelve biased simulations are performed 
with a fixed value of $\kappa = 0.05$~kJ/mol and different $N^*$-values that range from 4000 to 5100 in steps of 100. 
Each biased simulation is run for 1 ns, and the data from the first 400 ps are discarded as equilibration. 
For every biased simulation, $\Ntv$-values are stored every 50 time steps, 
whereas system configurations are saved every 500 time steps, and analyzed to obtain $H$.
System configurations are additionally used to obtain vapor-liquid interface profiles;
to this end, we average the water density field over 600 configurations,
and determine the iso-surface for which the averaged water density field is equal to half the bulk water density.
The vapor-liquid interface profile thus obtained is fitted to a circle, and the tangent to the fitted circle at the solid surface, i.e., at the $z$-coordinate corresponding to the outmost layer of solid atoms, is used to obtain an estimate the contact angle, which we call $\theta_{\rm I}$.
Error bars are estimated by dividing the production period into 3 to 5 blocks and determining the statistical uncertainty across the block averages.

\subsection{Droplet Simulations for Estimating Contact Angles}
To compare the wetting coefficients obtained using our approach with those obtained from the geometry of water droplets,
we also perform simulations of (infinitely long) cylindrical water droplets on the LJ surfaces described above. 
In these simulations, the solid surface is composed of 4200 LJ atoms, and has dimensions of 17.3607 nm $\times$ 2.8059 nm $\times$ 1.2887 nm. 
A box with dimensions of 17.3607 nm $\times$ 2.8059 nm $\times$ 10.0 nm is used to simulate 2000 water molecules that are placed on the surface.
The system is equilibrated for 4~ns, which is followed by a production period of 6~ns. 
In the production stage, configurations are collected every 1000 time steps, and are analyzed to characterize the shape of the droplet.
Droplet interfaces, the corresponding contact angles ($\theta_{\rm D}$), and associated error bars are obtained as described above.
%

\begin{figure*}[htb]
 \centering
 \includegraphics[width=0.8\textwidth]{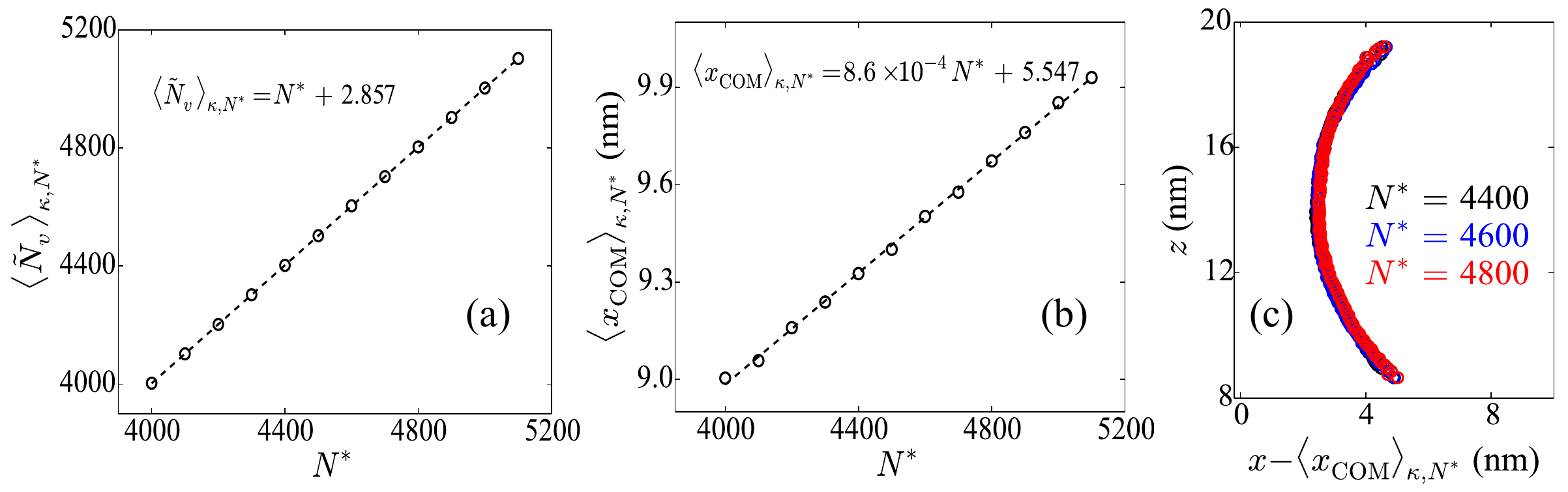}
 \caption{
Estimating the wetting coefficient for the LJ surface with $\esw = 1.94$~kJ/mol.
(a) The average number of coarse-grained waters in the observation volume, $\langle \Ntv \rangle_{\kappa,N^*}$ increases with $N^*$ (symbols).
Following Equation~\ref{eq:N}, the data are fit to a straight line with unit slope (dashed line), and the intercept is used to estimate $f$.
(b) The average $x$-position of the center of mass of the water slab, $\langle \zcom \rangle_{\kappa,N^*}$, also varies linearly with $N^*$ (symbols); a linear fit to the data (dashed line) is used to obtain the slope, $h$.
(c) Vapor-liquid interfaces, obtained from biased simulations with $N^*$-values of 4400, 4600 and 4800, 
are shifted in the $x$-direction 
by their respective $\langle \zcom \rangle_{\kappa,N^*}$-values.
The shifted interfacial profiles (obtained by averaging over 1000 configurations) agree well with one another, 
suggesting that the biasing potential influences the position of the interface, but not its shape.
}
 \label{fig1}
\end{figure*}

\section{Results and Discussions}
SWIPES is used to characterize the wetting coefficients ($k$) for LJ surfaces with wide range of solid-water attractions ($\esw$), and to estimate the vapor-liquid surface tension of water ($\gvl$).
The contact angles estimated using SWIPES ($\theta_{\rm F}$) are compared with those obtained using droplet geometries ($\theta_{\rm D}$). 
As $\esw$ is increased, a transition from partial to complete wetting is observed.
The computational efficiency of the method is also discussed.  

%
\subsection{Estimating wetting coefficients using SWIPES}
To illustrate our method, we first estimate $k$ for a surface with $\esw = 1.94$~kJ/mol. 
In Figure~\ref{fig1}a, we show the ensemble averages of the coarse-grained number of waters, $\langle \Ntv \rangle_{\kappa,N^*}$, evaluated at different $N^*$-values. 
The data are fit well by a straight line with unit slope (dashed line), confirming that $f$ is independent of $N^*$ (Equation~\ref{eq:N}).
The $y$-intercept of the fitted line corresponds to $-f/\kappa$, enabling estimation of $f = -0.143(3)$~kJ/mol.
As shown in the $\SI$, estimates of $f$ can also be obtained from individual biased simulations by using Equation~\ref{eq:f},
albeit with larger statistical uncertainties.
%
Figure~\ref{fig1}b shows that the position of vapor-liquid interface ($H$), quantified by $\langle \zcom \rangle_{\kappa,N^*}$, varies linearly with $N^*$; the corresponding slope is $h = 8.6(2)\times10^{-4}$~nm.
As shown in the $\SI$, $h$ can also be obtained from a single biased simulation by estimating the co-variance of $\zcom$ and $\Ntv$ (Equation~\ref{eq:covar}).
However, due to the larger statistical uncertainties associated with co-variance estimation,
the use of multiple biased simulations is likely to be more judicious route for estimating $h$. 
In Figure~\ref{fig1}c, vapor-liquid interfaces identified from three biased simulations are shown with their $x$-coordinates shifted by $\langle \zcom \rangle_{\kappa,N^*}$; the excellent agreement between the shifted interfacial profiles highlights the fact that biasing potential moves the water slab along the solid surface, but does not otherwise influence the shape of the interface.
Thus, Figure~\ref{fig1}c further corroborates the linear dependences of both $F_{\kappa,N^*}$ and $\langle \zcom \rangle_{\kappa,N^*}$ on $N^*$.
%
With the corresponding slopes $f$ and $h$ estimated, $k$ can then be obtained from Equation~\ref{eq:kdef2} if $\gvl$ is known. 
For the SPC/E water model used here, the surface tension has been estimated to be $\gamma_{\rm VL}^{\rm SPC/E} = 63.6(1.5)$~mN/m at 300~K~\cite{Vega}.
With this value of surface tension, we estimate the wetting coefficient 
for the LJ surface with $\esw = 1.94$~kJ/mol to be $k = 0.78(2)$, 
and the corresponding contact angle to be 38(2)$^{\circ}$. 

%
\subsection{Estimating vapor-liquid surface tension using SWIPES}
\begin{figure}[tb]
 \centering
 \includegraphics[width=0.34\textwidth]{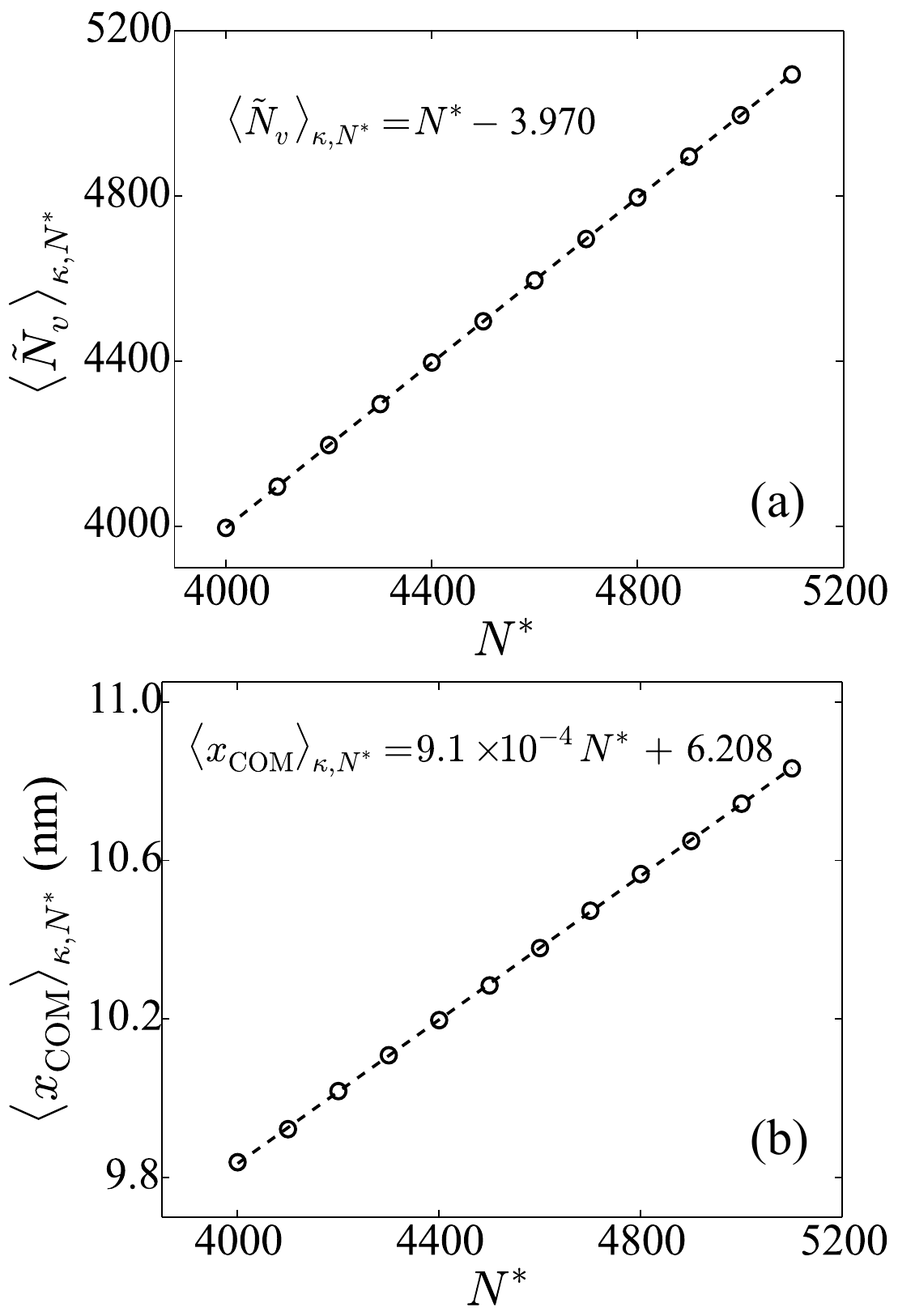}
 \caption{
Estimating the surface tension of SPC/E water using a non-wetting LJ surface with $\esw = 0.001$~kJ/mol.
(a) The increase in $\langle \Ntv \rangle_{\kappa,N^*}$ with $N^*$ (symbols), and a linear fit to the data (dashed line) are shown; in fitting the data, the slope is fixed at unity. 
(b) The variation of $\langle \zcom \rangle_{\kappa,N^*}$ with $N^*$ (symbols), and a linear fit to the data (dashed line) are shown.
}
 \label{fig2}
\end{figure}

In principle, the vapor-liquid surface tension, $\gvl$, of a fluid can be estimated using Equation~\ref{eq:kdef2} 
if we use SWIPES to characterize a surface (i.e., estimate $f$ and $h$) for which the wetting coefficient is known.
In practice, surfaces that are either fully wetting ($k=1$) or non-wetting ($k=-1$) would work well for this purpose.
However, purely repulsive surfaces are ideally suited for this purpose because they are non-wetting regardless of the fluid under consideration.
Here, we use an effectively repulsive hydrophobic surface with $\esw = 0.001$~kJ/mol to estimate $\gvl$ for SPC/E water.
For this surface, $\langle \Ntv \rangle_{\kappa,N^*}$ and $\langle \zcom \rangle_{\kappa,N^*}$ are shown as functions of $N^*$ in Figures~\ref{fig2}a and~\ref{fig2}b, respectively. 
Using this data, we obtain estimates of $f$ and $h$, which in turn enables us to estimate the vapor-liquid interfacial tension, 
$\gvl = (1/2L)\cdot(f / h) = 64(2)$~mN/m.
Our estimate is in excellent agreement with the value of 63.6~mN/m obtained in Ref.~\citenum{Vega} using the test-area method.

\begin{figure*}[htb]
\centering
 \includegraphics[width=0.7\textwidth]{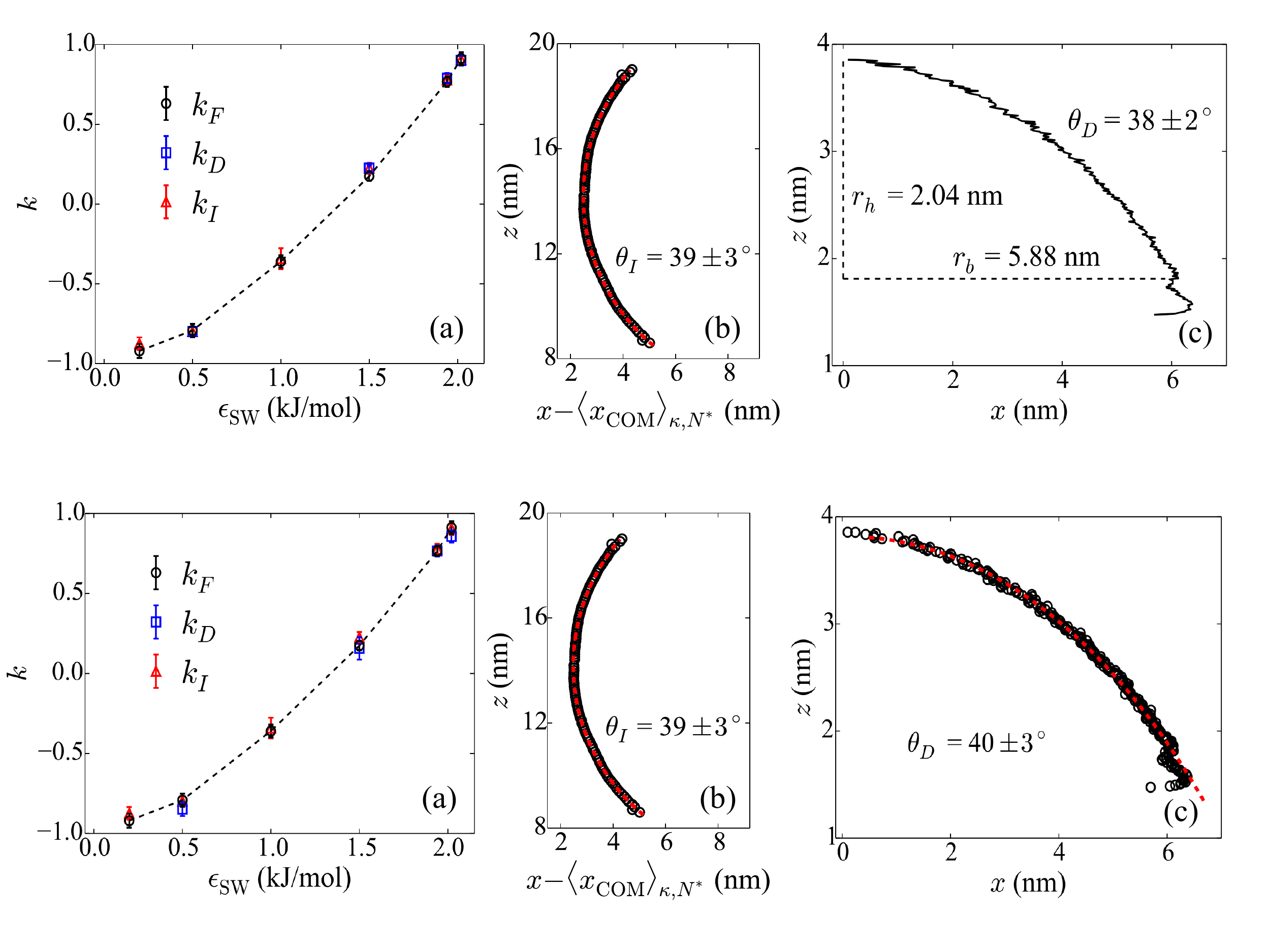}
  \caption{ 
(a) Wetting coefficients estimated using our free energy method ($k_{\rm F}$) are shown for LJ surfaces with a wide range of $\esw$-values.
These estimates are found to be consistent with the corresponding estimates obtained from the vapor-liquid interfacial profiles ($k_{\rm I} \equiv \cos\theta_{\rm I}$), 
as well as those obtained from separate simulations of cylindrical droplets ($k_{\rm D} \equiv \cos\theta_{\rm D}$). 
(b) For the LJ surface with $\esw = 1.94$~kJ/mol, the vapor-liquid interface obtained from a biased simulation with $N^*=4800$ is shown (symbols). 
The interface is fit to a circle (red dashed line) and the contact angle $\theta_{\rm I}$ is obtained from the tangent to the circle at 3-phase contact.
(c) The shape of a cylindrical droplet placed on an LJ surface with $\esw =1.94$~kJ/mol is shown. 
The droplet interface is fit to a circle (red dashed line) and the contact angle $\theta_{\rm D}$ is obtained from the tangent to the circle at 3-phase contact.
}
\label{fig3} 
\end{figure*}

\subsection{LJ surfaces with a wide range of surface-water attractions}
%
To highlight the versatility of SWIPES, we study surfaces with a wide range of solid-water interaction energies ($\esw$) 
that span an order of magnitude from roughly 0.2 to 2~kJ/mol.
In each case, linear dependences of $\langle \Ntv \rangle_{\kappa, N^*}$ and $\langle  \zcom \rangle_{\kappa, N^*}$ on $N^*$ 
are observed; see $\SI$.
The corresponding wetting coefficients are obtained using Equation~\ref{eq:kdef2},
and are shown as a function of $\esw$ in Figure~\ref{fig3}a. 
Over this range of $\esw$-values, $k$ increases monotonically with $\esw$, 
and varies from -0.92 to 0.91, nearly spanning the entire range of wetting coefficients possible. 

As highlighted in Figure~\ref{fig1}, the biasing potential wets the surface, but leave the curvature of the vapor-liquid interface inside $v$ unchanged.
Thus, the tangent of the vapor-liquid interface at 3-phase contact can also be used to estimate the contact angle from our biased simulations. 
For the surface with $\esw = 1.94$~kJ/mol, the contact angle is estimated to be $\theta_{\rm I} = 39(3)^{\circ}$ (Figure~\ref{fig3}b), 
and agrees well with the corresponding estimate $\theta_{\rm F} = 38(2)^{\circ}$ obtained using SWIPES.
%
%
To compare our estimates of $k$ with those obtained from the widely used sessile droplet method,
we additionally simulate water droplets placed on the solid surfaces, and estimate the corresponding contact angles. 
Although spherical droplets are typically employed for this purpose, the corresponding results tend to depend on droplet size due to the influence of line tension.
Following ref.~\cite{Cygan}, we thus make use of cylindrical droplets, which are infinitely long due to our use of periodic boundary conditions. 
For the surface with $\esw = 1.94$~kJ/mol, the droplet shape obtained from our simulations is shown in Figure~\ref{fig3}c,
and is used to estimate $\theta_{\rm D} = 40(3)^{\circ}$ following the procedure described in section C.
As shown in Figure~\ref{fig3}a, across surfaces that span a wide range of $\esw$-values, 
estimates of the wetting coefficients obtained from SWIPES ($k_{\rm F}$) 
are found to be consistent with those obtained from interfacial profiles ($k_{\rm I}$) and droplets ($k_{\rm D}$).
Vapor-liquid interface profiles from our biased simulations as well as 
geometries of cylindrical droplets for surfaces with other $\esw$-values are included in the $\SI$. 
%

%
Although the length of the 3-phase contact line remains constant in both the biased simulations and the simulations of cylindrical droplets,
the corresponding contact angles ($\theta_{\rm I}$ and $\theta_{\rm D}$) are expected to depend on the curvature of the vapor-liquid interface, 
which in turn depends on the separation between the two surfaces. 
In contrast, our estimates of $k_{\rm F}$ are not expected to be influenced by line tension or interface curvature effects, as discussed in section B.
The agreement between $k_{\rm F}$ and both $k_{\rm I}$ and $k_{\rm D}$ suggests that interface curvature are minimal (within the statistical uncertainty) for our choice of system size and inter-surface separation. 
As described in detail in the $\SI$, we studied two additional systems with smaller and larger separations, 
and find that $k_{\rm F}$ does not depend on separation, as expected.
We also estimate $k_{\rm I}$ for these systems, and find any differences across different inter-surface separation are within the statistical uncertainty.
%
Although estimates of $k_{\rm I}$ and $k_{\rm D}$ agree reasonably well with $k_{\rm F}$,
the necessity of having to choose a contact plane introduces an inherent ambiguity
in the determination of contact angles from geometric approaches~\cite{Rotenberg:JACS:2011,Sergi,Iglauer}. 
Such ambiguity is expected to introduce systematic errors in contact angle estimates obtained from geometry-based methods.
In contrast, the estimation of $k_{\rm F}$ is free from ambiguity, and uncertainties arise only from the sampling errors in estimating $\langle \Ntv \rangle_{\kappa, N^*}$ and $\langle \zcom \rangle_{\kappa, N^*}$. 
Thus, highly accurate estimates of surface wetting and interfacial properties can be obtained by simply increasing the number or length of the biased simulations, making SWIPES more suitable for studying small but important changes in wettability; e.g. changes in $k$ arising from changes in pressure, fluid composition or surface chemistry.
Moreover, the determination of contact angles from droplet geometry is challenging 
for both the most hydrophobic and the most hydrophilic surfaces. 
For surfaces with contact angles greater than $140^{\circ}$, the value of $\theta$ becomes increasingly sensitive to the choice of the contact plane, making its estimation from droplet geometry increasingly challenging for such hydrophobic surfaces.
Conversely, for hydrophilic surfaces with contact angles below $30^{\circ}$,
large system sizes and long equilibration times~\cite{Kanduc}
are needed for droplets to spread on the surface and achieve their equilibrium shapes.

\subsection{Fully Wetting and Dewetting Surfaces}
As solid-fluid interactions become more favorable,
$\theta$ decreases with the system eventually transitioning 
from partial wetting ($\theta>0^{\circ}$) to complete wetting ($\theta\to0^{\circ}$).
Understanding the factors that govern transitions between partial and complete wetting states is of interest in diverse contexts, including the design of materials, such as polymer nanocomposites~\cite{kumar2013nanocomposites,martin2015wetting,ferrier2016engineering}.
Conversely, as solid-fluid interactions are made increasingly unfavorable, the fluid is expected to dewet the solid.
Although the estimation of $\theta$ from droplet geometry becomes increasingly challenging for surfaces in the vicinity of wetting and dewetting transitions, here we illustrate that SWIPES nevertheless works well in those regimes.
%

\begin{figure}[tb]
 \centering
 \includegraphics[width=0.35\textwidth]{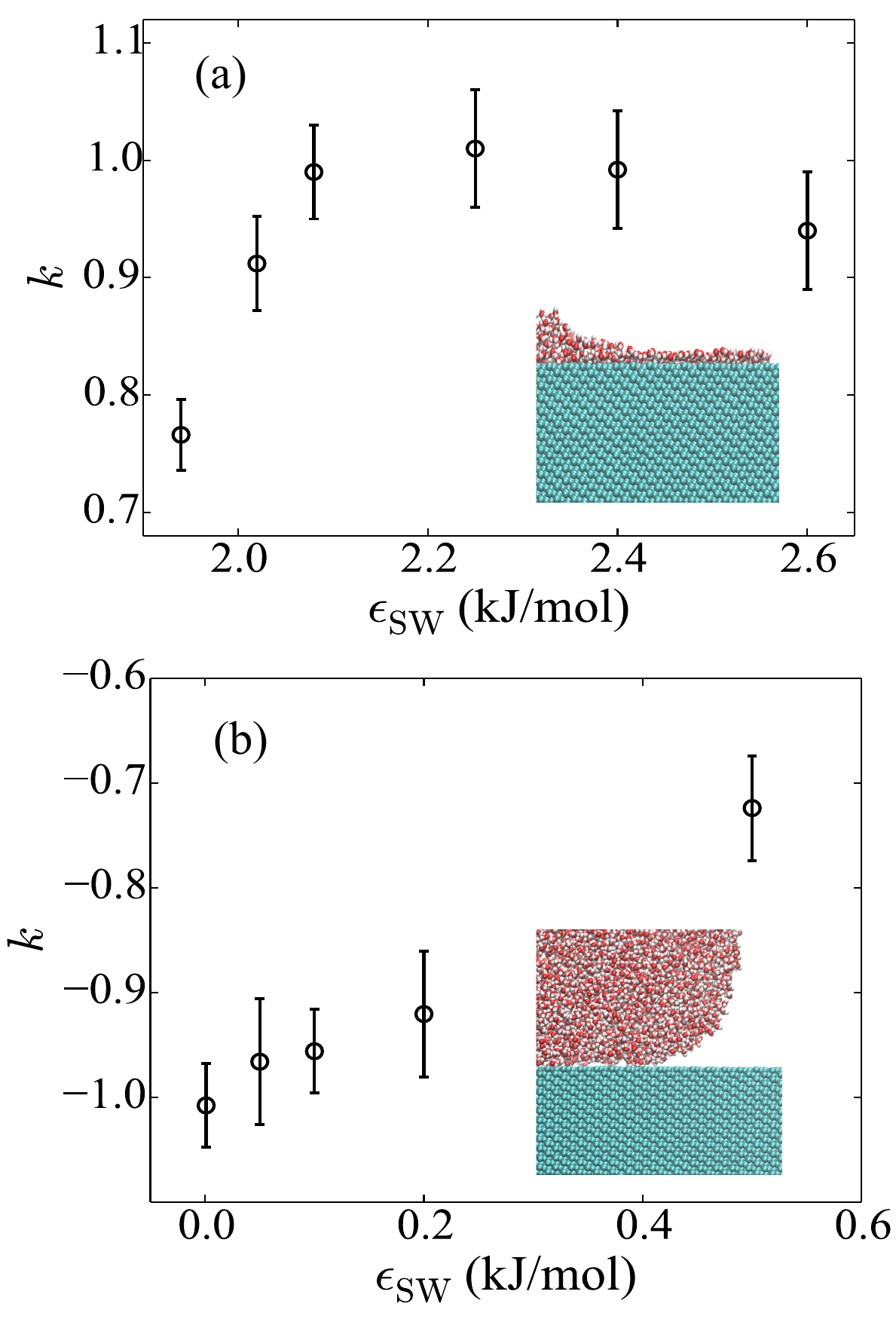}
  \caption{
(a) Wetting coefficients ($k$) are shown as a function of $\esw$ in the vicinity of the transition from partial to complete wetting. 
The inset is a simulation snapshot focusing on the liquid film that forms on the LJ surface with $\esw = 2.4$~kJ/mol. 
(b) Wetting coefficients ($k$) are shown as a function of $\esw$ near the transition from partial wetting to dewetting. 
The inset is a simulation snapshot focusing on the vapor layer that forms on the LJ surface with $\esw = 0.05$~kJ/mol. 
}
\label{fig4} 
\end{figure}

An LJ surface with a small value of $\esw$ interacts less favorably with liquid water than with its vapor phase, so that $ \gsv < \gsl$,
i.e., the surface is hydrophobic with $k = (\gsv - \gsl) /  \gvl < 0$.
As $\esw$ is increased, the surface interacts more favorably with both the liquid and vapor phases of water, 
i.e., both $\gsv$ and $\gsl$ decrease;
however, because the denser liquid can better utilize these favorable interactions, 
$\gsl$ decreases more rapidly than $\gsv$, and the wetting coefficient, $k = (\gsv - \gsl)/\gvl$ increases.
For a sufficiently large $\esw$-value, $\gsl$ eventually falls below $\gsv$, i.e., the surface becomes hydrophilic with $k = (\gsv - \gsl) /  \gvl > 0$.
Eventually, as $\esw$ is made large enough, any contact between solid and vapor is sufficiently unfavorable 
(relative to solid-liquid contact)
that $k \to 1$ and $\gsv \to \gsl + \gvl$.
In this limit, the free energetic penalty for replacing the solid-vapor interface 
with a solid-liquid interface and a vapor-liquid interface vanishes,
and water molecules from the vapor phase adsorb onto the surface to form a film of liquid.
As $\esw$ is increased even further, $\gsl$ does not continue to decrease more rapidly than $\gsv$;
rather the two quantities continue to decrease in concert with one another, so that $k$ plateaus at 1.
The phenomenology described above is generally true regardless of the solid and the fluid wetting it~\cite{binder2008modeling,Netz2}.
As shown in Figure~\ref{fig4}a, SWIPES captures such a transition from partial wetting ($k<1$) to complete wetting ($k=1$) as $\esw$ is increased, highlighting its suitability for characterizing highly hydrophilic surfaces.
The $\esw$-value at which the wetting coefficient, $k$ approaches 1 is roughly $\esw^* = 2.1$~kJ/mol,
with a further increase in $\esw$ leaving $k$ unchanged.
In the inset of in Figure~\ref{fig4}a, we show a simulation snapshot of an LJ surface with $\esw = 2.4$~kJ/mol ($>\esw^*$)
that is focused on the region near vapor-liquid-solid contact.
The snapshot highlights the presence of water molecules that are adsorbed onto the solid surface, 
and form a liquid film which covers the surface.
%

In addition to the transition to complete wetting, SWIPES also captures the transition from partial wetting to dewetting as $\esw$ is decreased. 
For small $\esw$-values, the surface is hydrophobic, and interacts more favorably with the vapor (relative to the liquid).
For sufficiently small $\esw$, any contact between solid and liquid is so unfavorable (relative to solid-vapor contact) 
that the surface dewets forming a vapor layer adjacent to the surface, in a manner that is analogous to the formation of a liquid film in the complete wetting regime.
The formation of such a vapor layer results in the solid-liquid interface being replaced with a solid-vapor interface and a vapor-liquid interface, and carries a vanishing free energetic penalty as $k \to -1$ and $\gsl \to \gsv + \gvl$.
As seen in Figure~\ref{fig4}b, the wetting coefficient indeed approaches -1 for small $\esw$.
In the inset of Figure~\ref{fig4}b, a simulation snapshot of a hydrophobic surface with $\esw = 0.05$~kJ/mol is shown; 
the region near the surface clearly shows the presence of a vapor layer.

An important feature of SWIPES is that both the liquid and vapor phases are always explicitly present by construction. 
This enables water molecules to adsorb onto the surface and form a liquid film in the complete wetting regime.
Similarly, the surface is readily able to dewet and form a vapor layer in the dewetting regime.
Moreover, because the vapor-liquid interface is moved along the surface in SWIPES,
as surface area comes in contact with liquid, it simultaneously loses contact with the vapor,
enabling direct estimation of $\gsv-\gsl$.
In contrast, methods that involve replacing the liquid adjacent to the surface with a cavity or a vacuum~\cite{Patel:PNAS:2011,Leroy}
instead estimate $\gamma_{\rm S,Vac} - \gsl$.
Because $\gamma_{\rm S,Vac}$ does not depend on $\esw$, and $\gsl$ decreases monotonically with increasing $\esw$, 
their difference, $\gamma_{\rm S,Vac} - \gsl$ does not plateau with increasing $\esw$.
Thus, such methods cannot identify the transition from partial to complete wetting unless $\gamma_{\rm S,Vac} - \gsv$ is estimated as well~\cite{Netz1}.

\begin{figure*}[htb]
\centering
\includegraphics[width=0.9\textwidth]{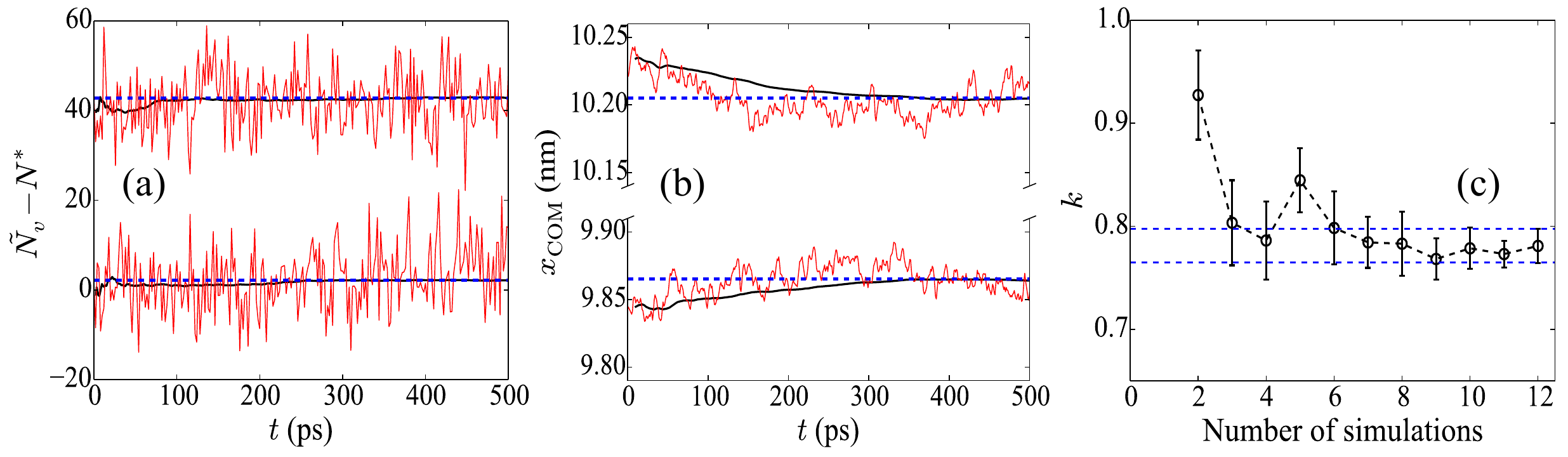}
  \caption{
For the LJ surface with $\esw$ = 1.94 kJ/mol, the time dependences of (a) $\Ntv - N^*$ and (b) $\zcom$ are shown for biased simulations with $N^*=4100$ and $N^*=4500$.
The values of $\Ntv - N^*$ have been shifted up by 40 for the $N^*=4500$ simulation for clarity.
Instantaneous values (red line), cumulative averages (black line) and ensemble averages (blue dashed line) highlight that convergence is achieved in 200 - 300~ps. 
(c) Wetting coefficients ($k$) calculated using different number of simulations. 
Blue dashed lines indicate the range of statistical uncertainty.
}
\label{fig5} 
\end{figure*}

\subsection{Computational Efficiency}
In addition to being relative simple, SWIPES is also efficient from a computational standpoint.
Here we provide a sense of the typical computational effort required to obtain accurate estimates of $f$ and $h$, and thereby of $k$. 
As described in section B, the estimation of $f$ relies on sampling of $\Ntv$,
and an accurate estimation of its average, $\langle \Ntv \rangle_{\kappa,N^*}$ (Equation~\ref{eq:f}). 
Figure~\ref{fig5}a shows the convergence of $\Ntv$ for two biased MD simulation with $N^*$ of 4100 and 4500 for the LJ surface with $\esw = 1.94$~kJ/mol. 
As shown in Figure~\ref{fig5}a, the cumulative average of $\Ntv - N^*$ (black line) converges to its equilibrium value (blue dashed line) in around 200~ps~\cite{Patel:JPCB:2014}. 
Similarly, Figure~\ref{fig5}b shows the convergence of $\zcom$ for the two biased simulations;
the corresponding cumulative average converges to its equilibrium value in roughly 300~ps. 

Although both $f$ and $h$ can also be estimated from individual biased simulations, 
the use of multiple simulations can yield estimates with lower statistical uncertainties.
For the LJ surface with $\esw = 1.94$~kJ/mol,
Figure~\ref{fig5}c shows wetting coefficient ($k$) estimates and uncertainties 
calculated using an increasing number of biased simulations;
every biased simulation has a production period of 600 ps, and the $N^*$-values of the biased simulations are separated by 100.
Reasonably accurate estimates of $k$ can be obtained with 3 to 6 biased simulations,
suggesting that a total simulation time of roughly 5 ns is sufficient.
This is comparable to the time needed to estimate contact angles from droplet shapes,
and roughly an order of magnitude smaller than other free energy methods for estimating $k$, 
such as the procedure developed by Kanduc and Netz~\cite{Netz2} or the dry-surface method~\cite{Leroy4}, 
which require 20 or more simulations, each run for 2~ns or more. 
We thus expect SWIPES to be comparable with or more efficient than existing methods from a computational standpoint.

\section{Conclusions and Outlook}
In this work, we introduce a method for the characterization of surface wetting and interfacial properties using enhanced sampling.
The method, which we call SWIPES, makes use of the indirect umbrella sampling (INDUS) technique to 
reversibly wet the surface of interest by biasing and systematically altering the coarse-grained number of fluid molecules 
in an observation volume $v$ adjacent to the surface.
SWIPES is versatile and can be used in conjunction with molecular dynamics simulations 
to estimate the wetting coefficient $k$ (or equivalently the contact angle, $\theta$) 
for systems with a wide range of solid-fluid interactions.
It can also be used to estimate the vapor-liquid surface tension of a fluid
by estimating the work required to wet a purely repulsive solvophobic surface.
Importantly, SWIPES involves displacing a vapor-liquid interface along the solid surface, but not the creation of additional vapor-liquid interfacial area.
In contrast, in a number of existing free energy methods, vapor-liquid interfaces spanning periodic boundaries are created;
the creation of such interfaces involves transitions between different dewetted morphologies, which can give rise to hysteresis, and complicate estimation of the corresponding free energies~\cite{Godawat,macdowell2006nucleation,grzelak2010nanoscale}.

We use SWIPES to characterize the wetting of a family of LJ surfaces by SPC/E water.
For partially wetting surfaces with a wide range of attractions for water,
our predictions for the contact angles are roughly consistent with those estimated from water droplet geometries.
As the attractions between the solid surface and water are increased, 
SWIPES is able to capture the transition from partial to complete wetting; 
it can also captures the dewetting transition at low attractions. 
Moreover, only ensemble averages of $\Ntv$ (number of fluid molecules in $v$) and $\zcom$ (center of mass of fluid in the direction perpendicular to the vapor-liquid interface) are required, making SWIPES both straightforward to implement and computationally efficient; a total simulation time of roughly 5~ns is sufficient to obtain reasonably accurate estimates of $k$. 

We note that although we have primarily discussed SWIPES in the context of surface-liquid-vapor systems, 
it can also be extended in a straightforward manner to other interfacial systems, such as surface-liquid-liquid or surface-liquid-solid systems.
By biasing an appropriate order parameter to control the extent to which a surface is covered by one phase relative to another,
the method presented here can be generalized to characterize the preference of a surface for one liquid over another, 
or for a crystalline solid over its coexisting liquid~\cite{fan2011amphiphilic,luu2014ellipsoidal,Errington5,Marks:PNAS:2018}.
We further note that our approach can also be used to estimate the free energetic cost for wetting surfaces with nanoscale heterogeneities, 
such as chemically patterned surfaces or surfaces with nanoscale texture;
for such surfaces, the notion of a contact angle itself may break down,
and even if it can be defined unambiguously, estimating the contact angle from droplet geometry may be non-trivial~\cite{Giovambattista:PNAS:2008,Acharya:Faraday:2010,Patel:JPCB:2012,harris2014effects,Xi:PNAS:2017,Wang19042011,Remsing:JPCB:2018}.
Another important class of problems for which the thermodynamic driving force can be characterized 
using our approach is the infiltration of fluids into porous media~\cite{huang2015polymer,shavit2015dynamics}.

\section*{Acknowledgements}
A.J.P. gratefully acknowledges financial support from the National Science Foundation (CBET 1511437, CBET 1652646, and UPENN MRSEC DMR 1720530), a grant from the Charles E. Kaufman Foundation (KA-2015-79204), and a fellowship from the Alfred P. Sloan Research Foundation (FG-2017-9406). 
H. J. was supported by a grant from the American Chemical Society's Petroleum Research Fund (56192-DNI6).
Z. V. was supported by a REACT fellowship through the National Science Foundation (OISE 1545884).



\balance




\bibliography{Wetting-INDUS} 
\bibliographystyle{rsc}

\end{document}


\section*{Supporting Information}
\label{secapenxa}\noindent{\Large\textbf{\textbf{Characterizing Surface Wetting and Interfacial Properties using Enhanced Sampling (SWIPES)}}}\\

\noindent Hao Jiang$^{\dag}$, Suruchi Fialoke $^{\dag}$, Zachariah Vicars $^{\dag}$ and Amish J. Patel$^{\dag}$\\
{\footnotesize\it $^{\dag}$Department of Chemical and Biomolecular Engineering, University of Pennslyvania, \\ 
\footnotesize\it Philadelphia, PA 19104, United States \\}

\clearpage

\section{Location of the observation volume $v$}
The observation volume is chosen to be a cuboid. 
The left edge of $v$ is placed around 1~nm to the left of the leftmost atoms in the solid surface.
As shown in Figure~1 of the main text, the observation volume covers the entire solid surface, and its right edge is placed to the right of the 
rightmost solid atoms.
Out of the two vapor-liquid interfaces that separate the liquid slab and vapor, the left interface remains well outside $v$ (far from the surface), whereas the right interface remains well within $v$ in all of our biased simulations.
These choices enable us to ensure that as the liquid slab moves along the surface, 
the only physical process that occurs is the replacement of vapor-solid interfacial area 
by an equivalent amount of liquid-solid interfacial area, and that there are no edge effects.

\section{Interface position $H$ and free energy $F_{\kappa,N^*}$ vary linearly with $N^*$}
To show that the interface position $H \equiv \langle \zcom \rangle_{\kappa,N^*}$ and free energy $F_{\kappa,N^*}$ in the biased ensemble vary linearly with $N^*$,
we first consider an ensemble with a constant number of coarse-grained waters, $\Nt$, in the observation volume, $v$.
The free energy difference between such an ensemble and the unbiased ensemble, $\fvnt$, 
is related to the statistics, $\pvnt = \langle \delta(\Ntv - \Nt) \rangle_0$, of coarse-grained water numbers in $v$ through $\beta \fvnt = -\ln \pvnt$; here $\langle \mathcal{O}(\Rbar) \rangle_0$ represents the average of $\mathcal{O}(\Rbar)$ in the unbiased ensemble.
In the constant-$\Nt$ ensemble, the position of the vapor-liquid interface, $H_n$, can be characterized using 
the corresponding ensemble average of the water slab center of mass position in the simulation box in the direction ($x$) perpendicular to the vapor-liquid interface, i.e., 
$H_n \equiv \langle \zcom \rangle_{\Nt}$, where $\langle \mathcal{O}(\Rbar) \rangle_{\Nt}$ represents the average of $\mathcal{O}(\Rbar)$ in the constant-$\Nt$ ensemble.

%
As the number of coarse-grained water molecules inside the observation volume ($v$) is increased by $\delta \Nt$,
the vapor-liquid interface ought to advance along the surface in $x$-direction by a distance $\delta H_n$.
However, the shape of the vapor-liquid interface is expected to be independent of $\Nt$;
instead, it is determined by the balance of forces at the 3-phase contact line,
which is governed entirely by the corresponding interfacial tensions.
Thus, $\delta \Nt$ and $\delta H_n$ ought to be related through:  
\begin{equation}
\delta \Nt = \delta H_n \times L \times \int_{z_{\rm low}}^{z_{\rm high}} [ \tilde\rho_{\rm L}(z) - \tilde\rho_{\rm V}(z) ] dz, 
\end{equation}
where $\tilde\rho_{\rm L}(z)$ and $\tilde\rho_{\rm V}(z)$ are the coarse-grained density profiles  (along the $z$-axis) of the liquid and vapor confined between the two solid surfaces, respectively.
$L$ is the length of the simulation box along the $y$-axis, and $z_{\rm low}$ and $z_{\rm high}$ are the lowest and highest coordinates of solid atoms that are in contact with water molecules, respectively. 
The corresponding slope, $h_n \equiv dH_n/d\Nt$ can then be expressed as:
\begin{equation}
h_n = \frac{1}{L\times \int_{z_{\rm low}}^{z_{\rm high}} [ \tilde\rho_{\rm L}(z) - \tilde\rho_{\rm V}(z) ] dz}.
\end{equation}
For a homogeneous flat solid surface, the water density profile remains unchanged regardless the position of the interface, and both $z_{\rm low}$ and $z_{\rm high}$ are constants. Therefore, the slope $h_n$ is a constant, i.e. $H_n$ is a linear function of $\Nt$. 
%
Because free energy and interface position are linearly related to one another (Equation 3 of the main text),
$\fvnt$ should also be a linear function of $\Nt$, making $f_n \equiv d\fvnt/d\Nt$ a constant.

Moreover, the free energetics of the constant-$\Nt$ ensemble, $\fvnt$, and the free energetics of biased ensemble, $F_{\kappa,N^*}$, are related through~\cite{Xi1, Xi2}
\begin{equation}
F_v(\Nt) = F_{v}^{\kappa,N^*}(\Nt) - U_{\kappa,N^*}(\Nt) + F_{\kappa,N^*},
\label{eq:us} 
\end{equation}
where $U_{\kappa,N^*}(\Nt) = \frac{\kappa}{2} ( \Nt - N^* )^2$ is the parabolic biasing potential,  
and $\beta F_v^{\kappa,N^*} (\Nt) = - \ln P_v^{\kappa,N^*}(\Nt)$ represents the free energetics of $\Ntv$ in the biased ensemble, 
with the corresponding statistics being given by $P_v^{\kappa,N^*}(\Nt) = \langle \delta(\Ntv - \Nt) \rangle_{\kappa,N^*}$.
%
Taking the derivative of Equation~\ref{eq:us} with respect to $\Nt$:
\begin{equation}
f_n \equiv \frac{\partial F_v}{\partial \Nt} = \frac{\partial F_v^{\kappa,N^*} }{ \partial \Nt} - \kappa (\Nt - N^*),
\label{eq:usd} 
\end{equation}
where as discussed above, $f_n$ is a constant (independent of $\Nt$). 
By recognizing that $F_v^{\kappa,N^*}(\Nt)$ ought to have a minimum at $\Nt = \langle \Ntv \rangle_{\kappa,N^*}$, 
and that the corresponding $\Nt$-derivative ought to be 0, we get:
\begin{equation}
f_n = - \kappa [ \langle \Ntv \rangle_{\kappa,N^*} - N^* ]. 
\end{equation}
From Equation~6 of the main text, we then see that 
\begin{equation}
f_n = \kappa [ N^* - \langle \Ntv \rangle_{\kappa,N^*} ] = f. 
\end{equation}
Because $f_n$ is a constant, $f \equiv dF_{\kappa,N^*}/dN^*$ should also be a constant.
Thus, $F_{\kappa,N^*}$ must be a linear function of $N^*$.
Finally, because the free energetics, $F_{\kappa,N^*}$, and interface position, $H \equiv \langle \zcom \rangle_{\kappa,N^*}$, in the biased ensemble are linearly related to one another (Equation 3 of the main text), $H$ must also a linear function of $N^*$, and $h$ a constant.



For the LJ surface with $\epsilon_{\rm SW}$=1.94 kJ/mol, we used umbrella sampling to obtain $\fvnt$;
i.e., we employed biased simulations with sufficient overlap in the distributions, $P_v^{\kappa,N^*}(\Nt)$, between neighboring windows.
We then analyzed the results using standard techniques, such as the weighted histogram analysis algorithm (WHAM)~\cite{Souaille}
to obtain both $\fvnt$ as a function of $\Nt$, and $F_{\kappa,N^*}$ as a function of $N^*$. 
These results are shown in Figures~S1a and~S1b,
%
and confirm the linear dependences of $\fvnt$ on $\Nt$ and of $F_{\kappa,N^*}$ on $N^*$;
the corresponding slopes, $f$ and $f_n$ also agree with one another, as expected.
The simple (linear) dependences of $\fvn$ and $F_{\kappa,N^*}$ as well as of $H_n$ and $H$ 
makes the estimation of $f$ and $h$ both straightforward and efficient,
and lie at the heart of the simplicity and computational efficiency of SWIPES.

\begin{figure}[H]
   \centering
    \includegraphics[height=5cm]{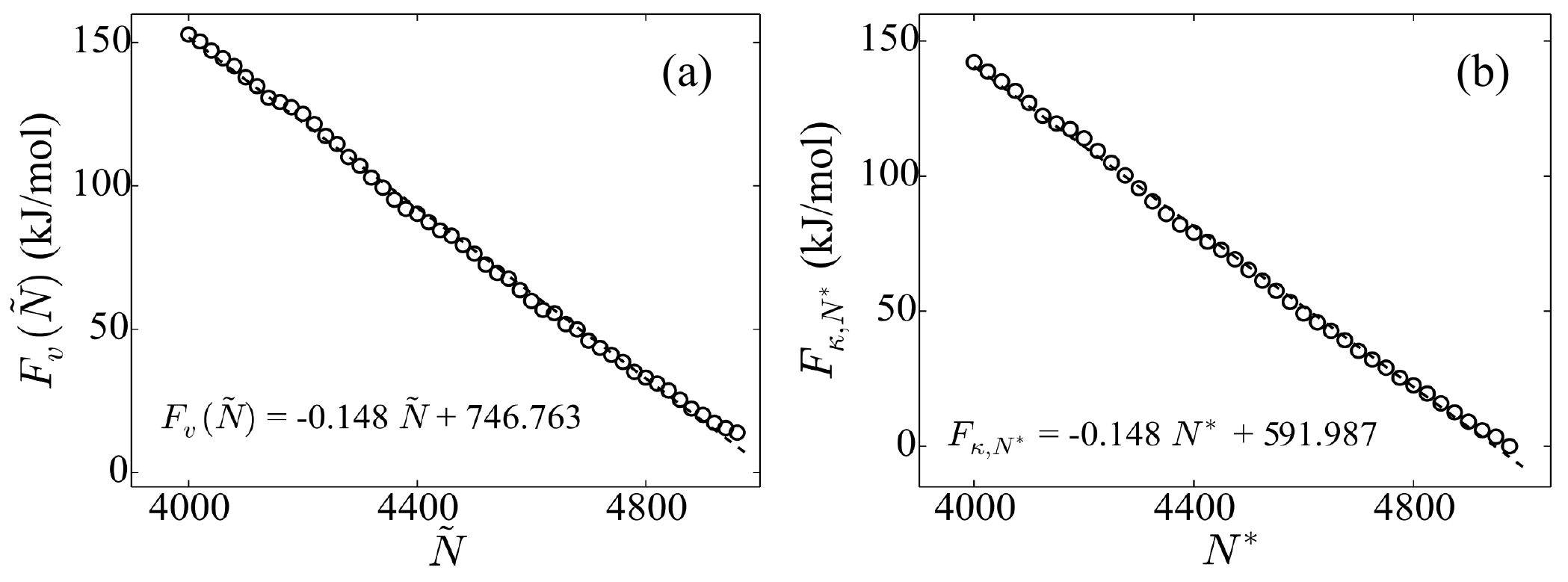} 
  \caption{
For the LJ surface with $\epsilon_{\rm SW} =1.94$~kJ/mol: 
(a) $\fvnt$ calculated using umbrella sampling and WHAM (black symbols) is shown as a function of $\Nt$.
%
(b) $F_{\kappa,N^*}$ calculated using umbrella sampling and WHAM (black symbols) is shown as a function of $N^*$.
%
The dashed lines are linear fits to the data.
}
\label{fig1} 
\end{figure}

\section{Estimating $f$ from individual biased simulations}
By using Equation 7 of the main text, $f$ can be estimated from every biased simulation.
Such estimates are shown in Figure~S2 (symbols) for the LJ surface with $\epsilon_{\rm SW} = 1.94$~kJ/mol. 
The average of $f$ over the 12 independent simulations is indicated by the horizontal dashed line, 
and is in agreement with $f$ calculated from the $y$-intercept of $\langle N_v \rangle_{\kappa,N^*}$ vs $N^*$; 
see Figure~2a of the main text.

\begin{figure}[H]
   \centering
    \includegraphics[height=5cm]{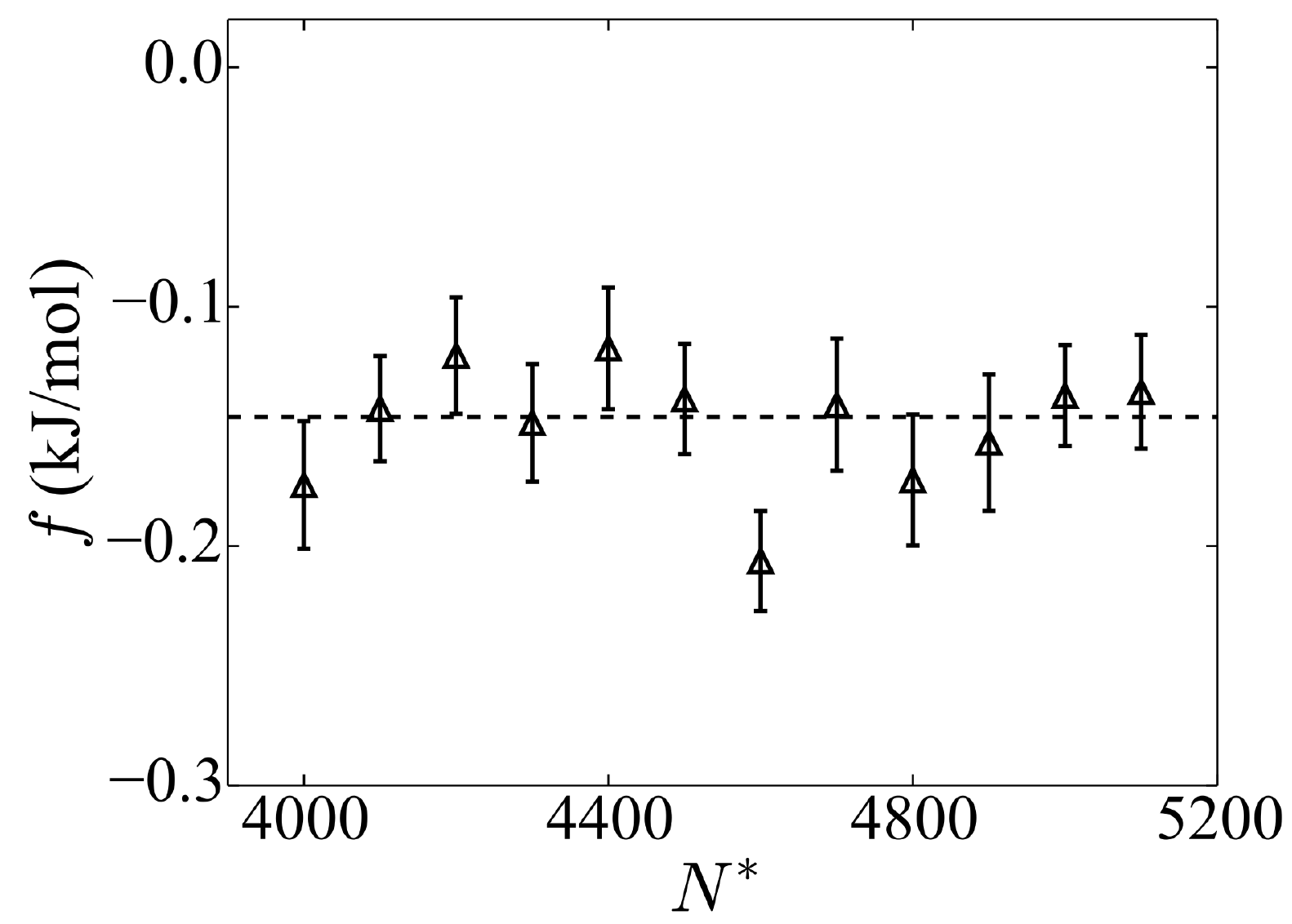} 
  \caption{
For the LJ surface with $\epsilon_{\rm SW} = 1.94$~kJ/mol, $f = \kappa (N^* - \langle \tilde{N_v}\rangle_{\kappa,N^*})$ (Equation 7 of the main text) is estimated from each biased simulation (symbols).
The horizontal dashed line is the value of $f$ averaged over the 12 $f$-estimates obtained from the biased simulations. 
}
\label{fig1} 
\end{figure}

\section{Estimating $h$ from covariances}
From every biased simulation, the slope, $h \equiv d \langle x_{\rm COM} \rangle_{\kappa,N^*} / d N^*$  
can be estimated using the co-variance of the fluctuations in $x_{\rm COM}$ and $\tilde{N_v}$ (Equation 9 of the main text).
For the LJ surface with $\epsilon_{\rm SW} = 1.94$~kJ/mol, $h$ thus obtained is shown as a function of $N^*$ in Figure~S3;
it is clear that the statistical uncertainties associated with the corresponding estimates of $h$ is substantial.
The average of the $h$-estimates obtained using the co-variance relation from the 12 biased simulations is $7.8(\pm1)\times10^{-4}$~nm  (dashed line in Figure~S3), and agrees reasonably well with the 
estimate of $h$ obtained from the slope of the line fitted to $\langle x_{\rm COM} \rangle_{\kappa,N^*}$ vs $N^*$ 
(Figure~2b of the main text). 
\\

\begin{figure}[H]
   \centering
    \includegraphics[height=5cm]{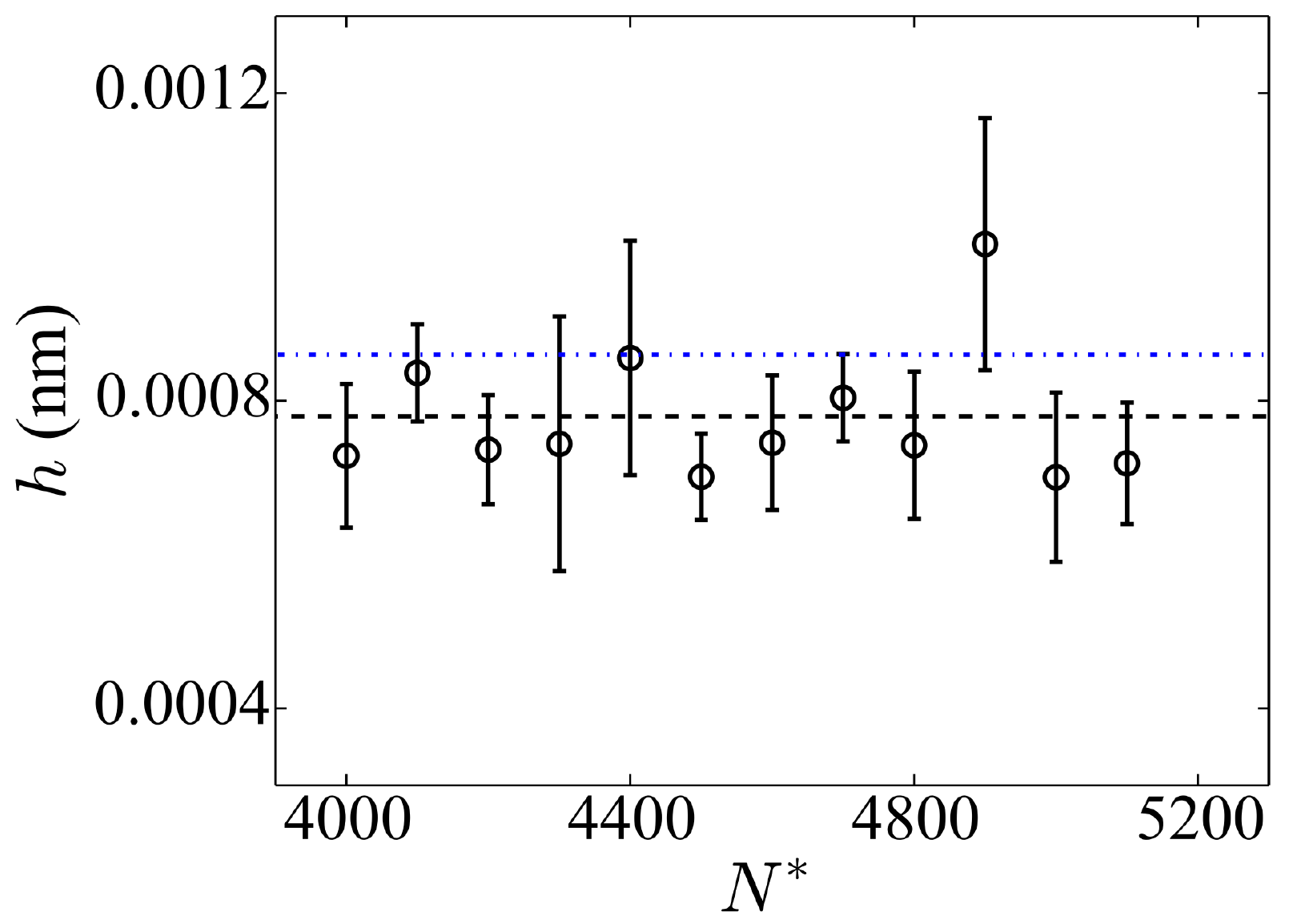} 
  \caption{
  For the LJ surface with $\epsilon_{\rm SW} = 1.94$~kJ/mol, $h$ estimated from the co-variance of $x_{\rm COM}$ and $\tilde{N_v}$ is shown for each biased simulation (symbols). The dashed line (black) indicates the average of the 12 $h$-estimates obtained from the biased simulations. The dot-dashed line (blue) indicates the value of $h$ obtained in Figure~2b of the main text. }
\label{fig1} 
\end{figure}

\section{$\langle \tilde{N_v} \rangle_{\kappa,N^*}$ and $\langle x_{COM} \rangle_{\kappa,N^*}$ for LJ surfaces with different $\epsilon_{\rm SW}$-values}

\begin{figure}[H]
\centering
\includegraphics[width=\textwidth]{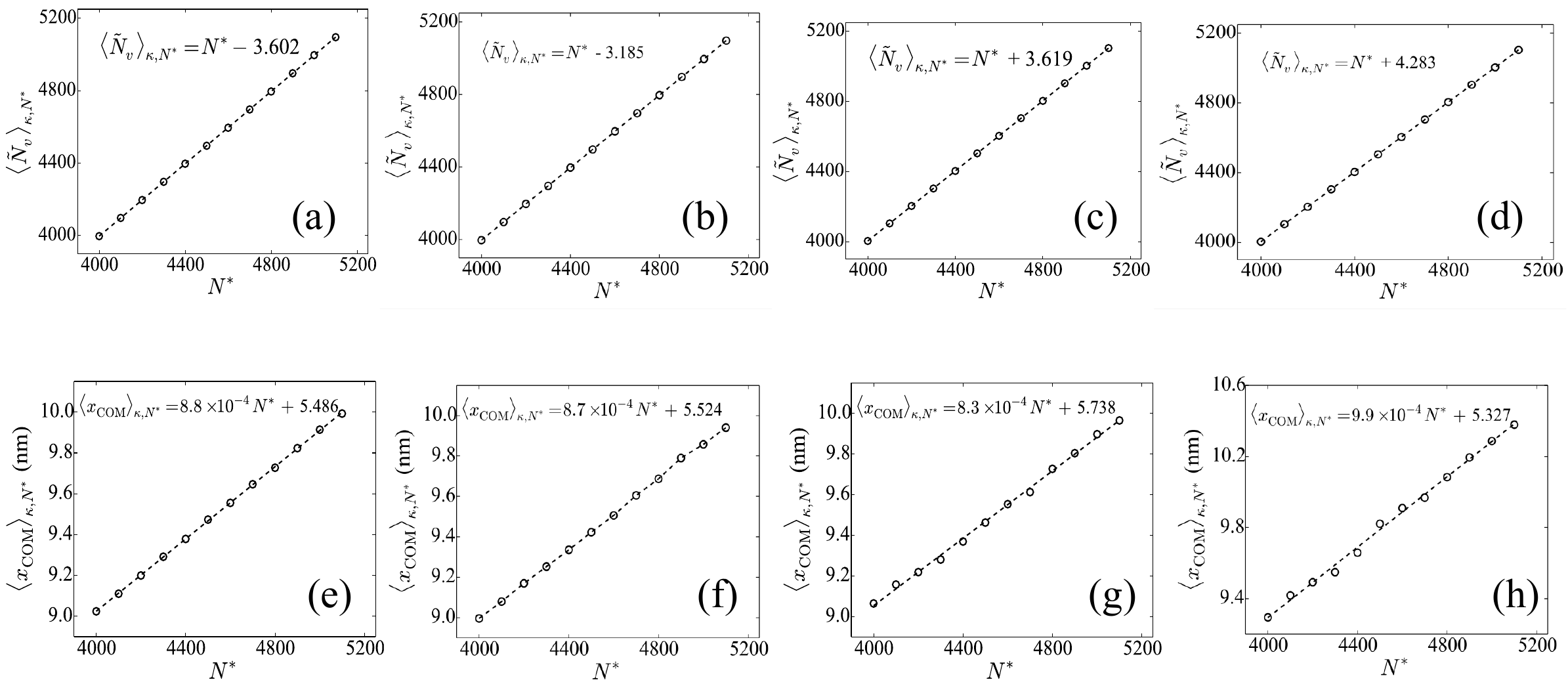} 
  \caption{
Ensemble averages of the coarse-grained number of water in the observation volume, $\langle\tilde{N_v}\rangle_{\kappa,N^*}$, are shown as a function of $N^*$ for LJ surfaces with $\epsilon_{\rm SW}$-values of (a) 0.1 kJ/mol, (b) 0.5 kJ/mol, (c) 2.08 kJ/mol, and (d) 2.25 kJ/mol, respectively. 
Ensemble averages of water slab center of mass position along the $x$-axis, $\langle x_{\rm COM} \rangle_{\kappa,N^*}$, are shown as a function of $N^*$ for LJ surface with $\epsilon_{\rm SW}$-values of (e) 0.1 kJ/mol, (f) 0.5 kJ/mol, (g) 2.08 kJ/mol, and (h) 2.25 kJ/mol, respectively. 
Dashed lines are linear fits to the simulation data (symbols). 
In fitting $\langle \tilde{N_v}\rangle_{\kappa,N^*}$ vs. $N^*$, the slope is set to 1.  
}
\label{fig1} 
\end{figure}

\section{Effect of solid surfaces separation distance on the calculation of $k$}
For all the biased simulations described in the main text, the length of the simulation box in the $z$-direction, $L_z$, (which sets the separation between the solid surface) was fixed to be 14.319~nm.
Here we estimate $k$ using a smaller and larger systems with $L_z$-values of 10.023 nm and 17.182 nm, respectively.
%
While 7000 water molecules were used in the biased simulations reported in the main text, 5000 and 12000 waters are used in the small and large systems, respectively.  
%
The solid surfaces themselves continue to be made up from 8640 atoms. 
In Figure~S5, the ensemble averages $\langle\tilde{N_v}\rangle_{\kappa,N^*}$ and $\langle x_{\rm COM} \rangle_{\kappa,N^*}$ are shown as functions of $N^*$ for the small and large systems, and enable estimation of $k$.
The contact angles thus estimated from the small and large systems using SWIPES are 40(2)$^{\circ}$ and 41(2)$^{\circ}$, respectively,
and agree with one another as well as with the system used in the main text within statistical uncertainty.  
Alternatively, contact angles can also be extracted from vapor-liquid interface geometries, as shown in Figure~S6. 
The contact angles thus estimated for the small and large systems are 38(3)$^{\circ}$ and 39(3)$^{\circ}$, respectively, 
and agree with one another as well as with the system used in the main text within statistical uncertainty.

\begin{figure}[H]
\centering
\includegraphics[width=0.8\textwidth]{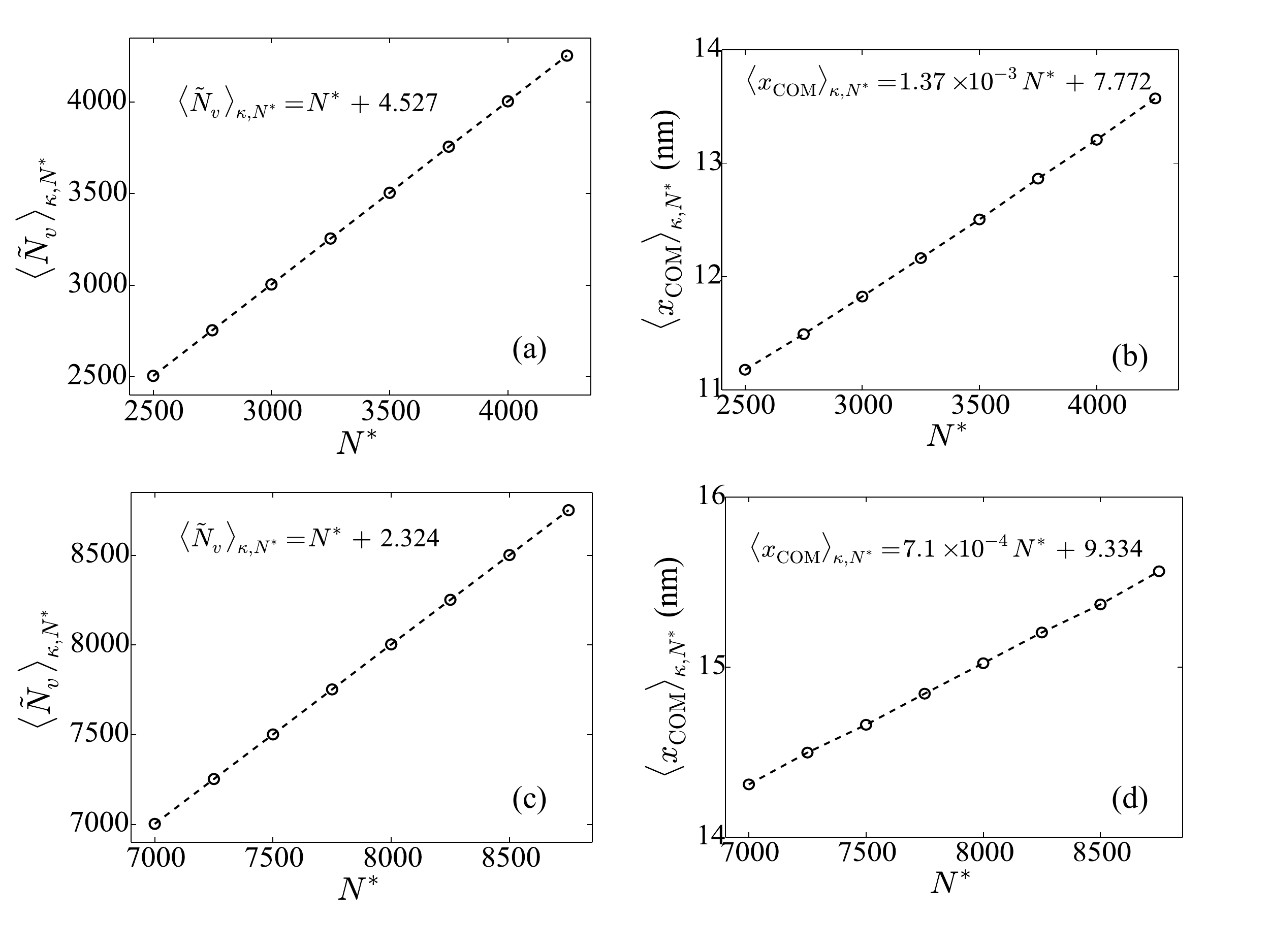} 
  \caption{
For the LJ surface with $\epsilon_{\rm SW} = 1.94$~kJ/mol: (a) Ensemble average of the coarse-grained number of waters in the observation volume, $\langle\tilde{N_v}\rangle_{\kappa,N^*}$, as a function of $N^*$ obtained from biased simulations using the small system (see text). 
(b) Ensemble average of water slab center of mass position along the x-axis, $\langle x_{\rm COM} \rangle_{\kappa,N^*}$, as a function of $N^*$ obtained from biased simulations using the small system. 
(c) Ensemble average of the coarse-grained number of water in the observation volume, $\langle\tilde{N_v}\rangle_{\kappa,N^*}$, as a function of $N^*$ obtained from biased simulations using the large system (see text). 
(d) Ensemble average of water slab center of mass position along the x-axis, $\langle x_{\rm COM} \rangle_{\kappa,N^*}$, as a function of $N^*$ obtained from biased simulations using the large system.
}
\label{fig1} 
\end{figure}

\begin{figure}[H]
\centering
\includegraphics[height=6cm]{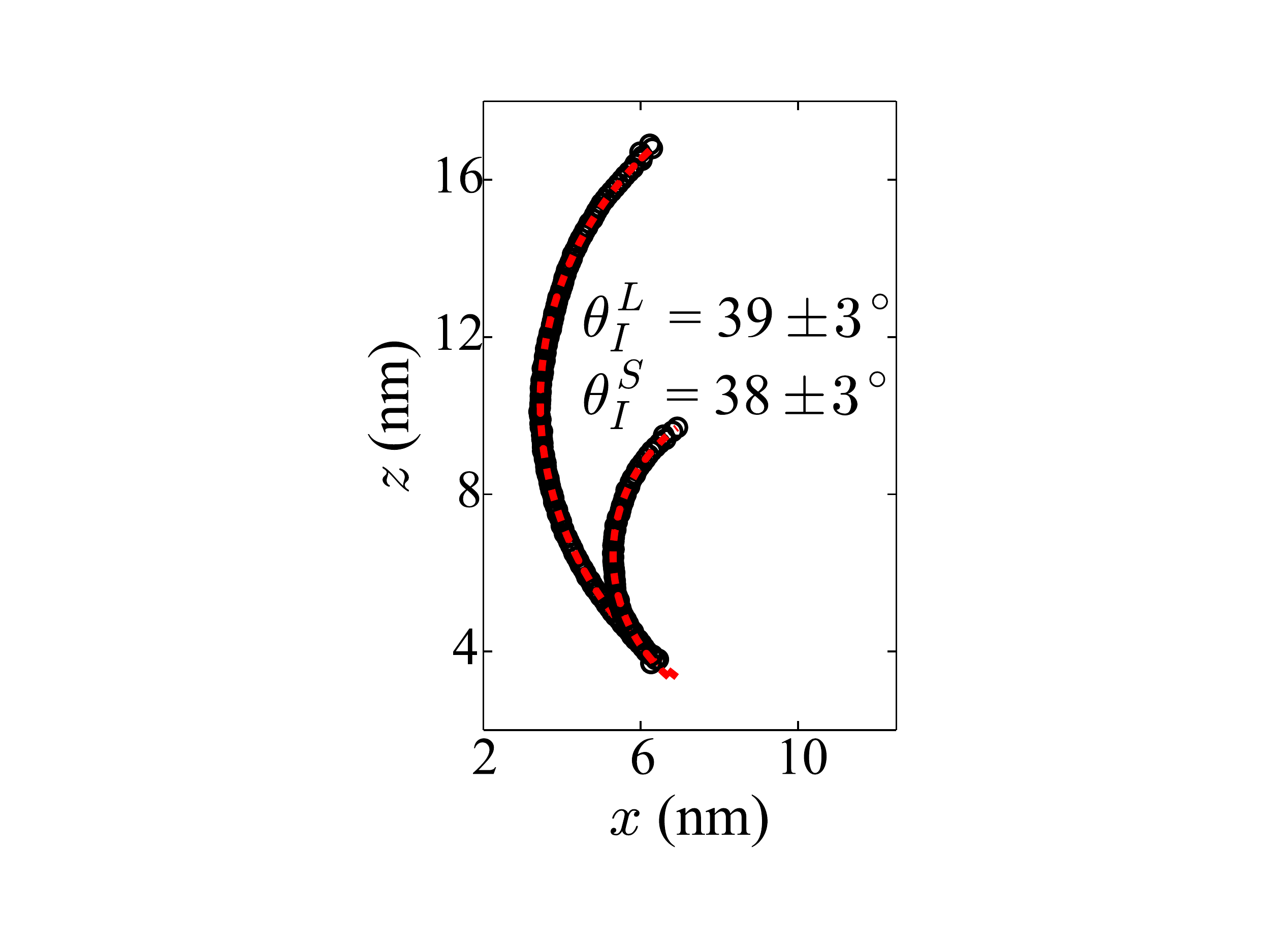} 
  \caption{Vapor-liquid interface profiles for the LJ surface with $\epsilon_{\rm SW}$ =1.94 kJ/mol obtained from biased simulations of the large ($L$) and small ($S$) systems; the interface profiles are fit to circles (red dashed lines). }
\label{fig1} 
\end{figure}

\section{Geometry of vapor-liquid interfaces}
\begin{figure}[H]
   \centering
    \includegraphics[height=6cm]{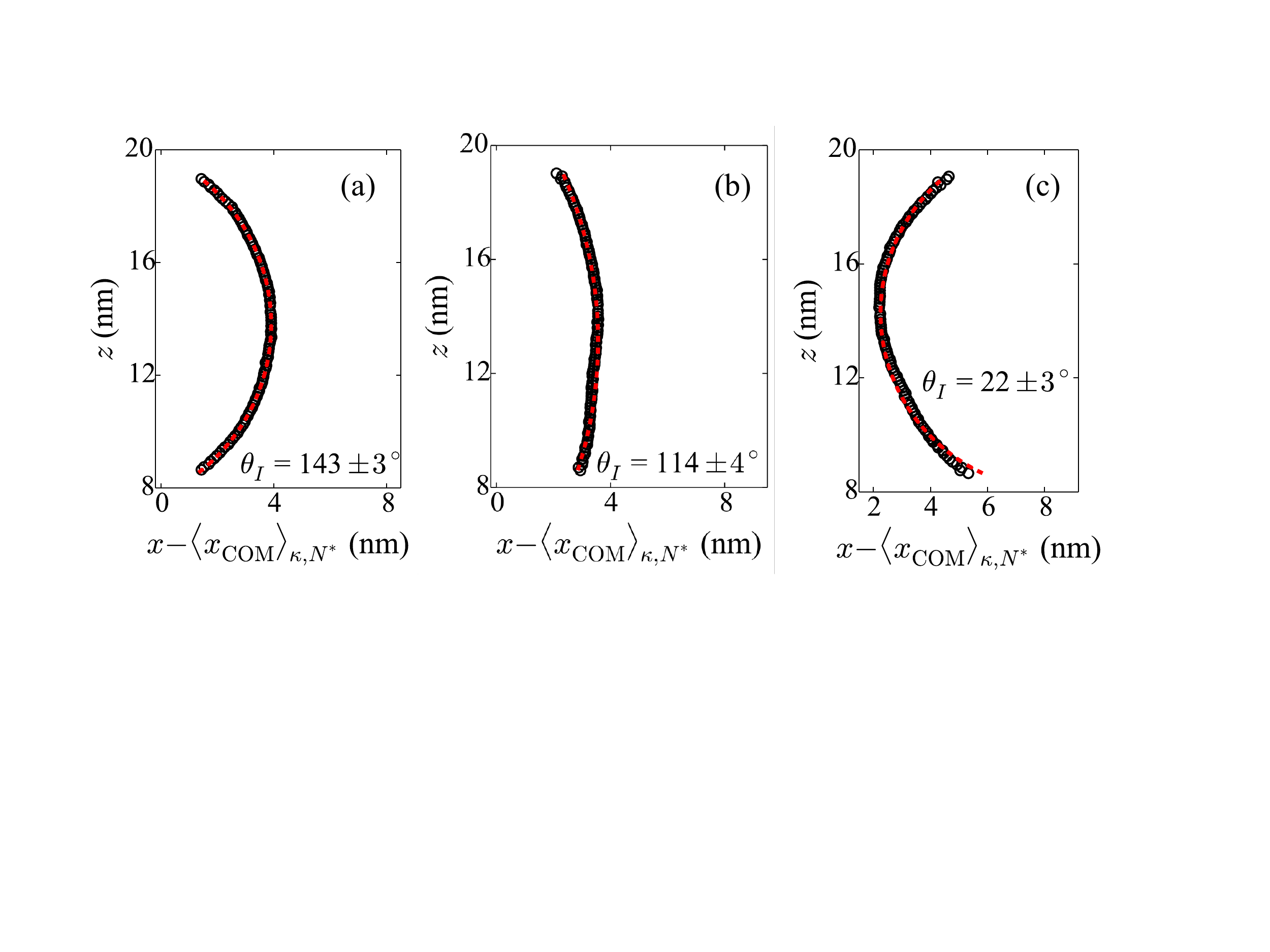} 
  \caption{
  Vapor-liquid interface profiles obtained from our biased simulations for LJ surfaces with $\epsilon_{\rm SW}$ of (a) 0.5 kJ/mol, (b) 1.0 kJ/mol, and (c) 2.02 kJ/mol, respectively; the profiles are fit to circles (red dashed lines).
  }
\label{fig1} 
\end{figure}

\section{Geometry of cylindrical droplets}
\begin{figure}[H]
   \centering
    \includegraphics[height=5cm]{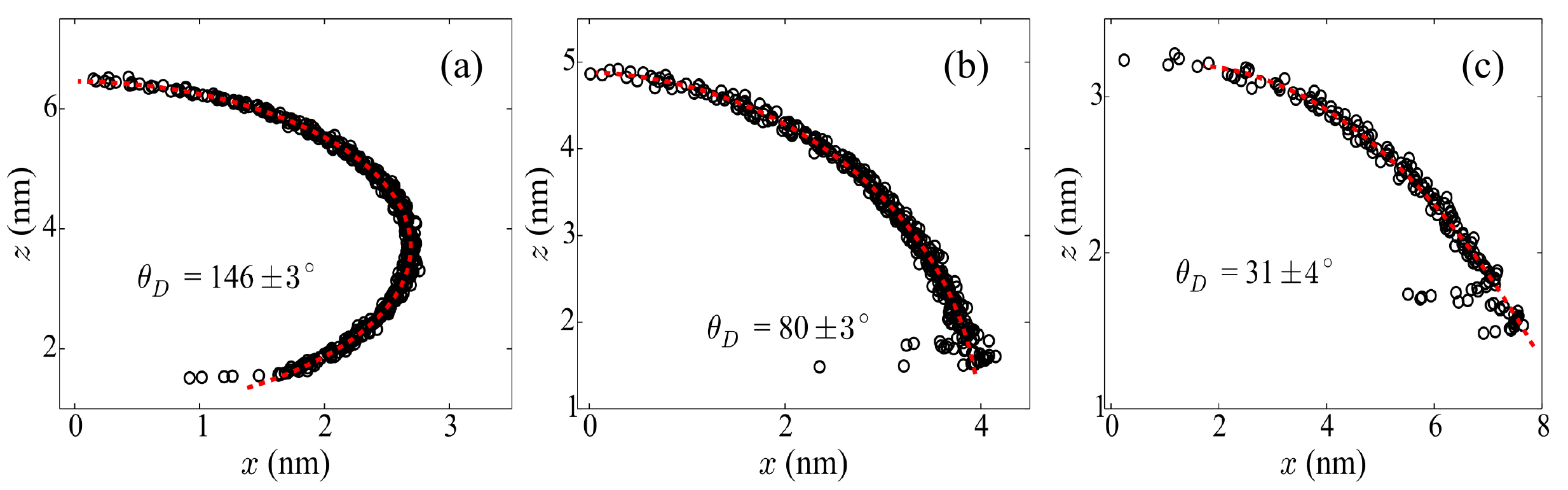} 
  \caption{Vapor-liquid interface profiles for cylindrical water droplets on LJ surfaces with $\epsilon_{\rm SW}$ of (a) 0.5 kJ/mol, (b) 1.5 kJ/mol, and (c) 2.02 kJ/mol, respectively; the profiles are fit to circles (red dashed lines).
  }
\label{fig1} 
\end{figure}

\footnotesize{

\bibliography{../Wetting-INDUS} 
\bibliographystyle{rsc} 

}